\documentclass[10pt]{cicp}

\usepackage[legalpaper, portrait, margin=2cm]{geometry}

\usepackage{bm}
\usepackage{color}
\usepackage{amsmath, amssymb}
\usepackage{ulem}
\usepackage{graphicx}
\usepackage{placeins}
\usepackage{algorithm}
\usepackage{algorithmic}
\usepackage{url}
\usepackage{subcaption}

\definecolor{purple}{rgb}{0.6, 0.4, 0.8}

\begin{document}

\title{Non-uniqueness of the Quasinormal Mode Expansion of Electromagnetic Lorentz Dispersive Materials}
\author[A.~Gras et.~al]{Alexandre Gras\affil{1,2}, Philippe Lalanne\affil{2}, and Marc Durufl\'e\affil{1}\comma\corrauth}
\address{\affilnum{1}\ INRIA Bordeaux Sud-Ouest, Talence 33405, France. \\
  \affilnum{2}\ LP2N, Institut d'Optique Graduate School, CNRS, Univ. Bordeaux, 33400 Talence, France}

\emails{{\tt marc.durufle@inria.fr} (M.~Durufl\'e), {\tt alexandre.gras@institutoptique.fr} (A.~Gras), {\tt Philippe.Lalanne@institutoptique.fr} (P.~Lalanne)}

\providecommand{\keywords}[1]
{
  \small	
  \textbf{\textit{Keywords---}} #1
}

\begin{abstract}
Any optical structure possesses resonance modes and its response to an excitation can be decomposed onto the quasinormal and numerical modes of discretized Maxwell's operator. In this paper, we consider a dielectric permittivity that is a N-pole Lorentz function of the pulsation $\omega$. We propose a common formalism and obtain different formulas for the modal expansion. The non-uniqueness of the excitation coefficient is due to a choice of the linearization of Maxwell's equation with respect to $\omega$ and of the form of the source term. We make the link between the numerical discrete modal expansion and analytical formulas that can be found in the literature. We detail the formulation of dispersive Perfectly Matched Layers (PML) in order to keep a linear eigenvalue problem. We also give an algorithm to regain an orthogonal basis for degenerate modes. Numerical results validate the different formulas and compare their accuracy. 
%
\end{abstract}


\maketitle

\section{Introduction}

\subsection{Quasinormal Modes}

Optical micro and nanoresonators, be they plasmonic, photonic or hybrid, enhance and localize the electromagnetic energy at wavelength or subwavelength scales and are key components in many photonic applications. Their optical response is characterized by one of a few resonant features resulting from the excitation of one or a few dominant modes, the natural resonance modes of the resonators.
These modes conveniently labelled by the integer $m=1,2 ...$ are characterized by their electric and magnetic field vectors distributions, $\widetilde{\textbf{E}}_m(\textbf{r})$ 
and $\widetilde{\textbf{H}}_m(\textbf{r})$. These vectors are solutions of the following eigenvalue boundary problem \cite{LalanneReview} 
\begin{equation}
\left \{ 
\begin{array}{lll}
-i \, \tilde{\omega}_m \varepsilon(\tilde{\omega}_m) \widetilde{\textbf{E}}_m - \nabla \times \widetilde{\textbf{H}}_m & = & 0, \medskip \\
-i \, \tilde{\omega}_m \mu(\tilde{\omega}_m) \widetilde{\textbf{H}}_m + \nabla \times \widetilde{\textbf{E}}_m & = & 0, \medskip \\ 
+ \mbox{ Boundary conditions} & &
\end{array}
\right. 
\label{eq:EigenvalueProb}
\end{equation}
where $\varepsilon(\tilde{\omega}_m)$ and $\mu(\tilde{\omega}_m)$ are respectively the dielectric permittivity and magnetic permeability and depend on the position $\text\bf{r}$ and pulsation $\omega$.
The fields $\widetilde{\textbf{E}}_m(\textbf{r})$ have continuous tangent traces across interfaces between subdomains and satisfy the outgoing-wave conditions at infinity. The $\exp(-i\omega t)$ convention for time harmonic fields is assumed. They are often called quasinormal modes (QNMs) to emphasize that their harmonic evolution is characterized by an exponential damping in time (they are the eigenstates of a non-Hermitian operator), so to say their pulsation $\tilde{\omega}_m$ is complex with Im$(\tilde{\omega}_m)<0$. 
Micro and nanoresonators play a leading role in many areas in nanophotonics, from quantum information processing to ultrasensitive biosensing, nonlinear optics, and various optical metasurfaces. This pushes a strong pressure on the development of QNM theory and QNM numerical methods that explicitly consider QNMs in the analysis, providing important clues towards the interpretation of the resonator response.

\subsection{Quasinormal Mode expansion of the scattered field}

The scattered field $\left[ \textbf{E}_S(\textbf{r},\omega), \textbf{H}_S(\textbf{r},\omega)\right]$ is solution of time-harmonic Maxwell's equations
\begin{equation*}
\left \{ 
\begin{array}{lll}
-i \, \omega \, \varepsilon(\omega) \textbf{E}_S - \nabla \times \textbf{H}_S & = & i \omega (\varepsilon(\omega) - \varepsilon_b) \, \textbf{E}_{\mbox{inc}}, \medskip \\
-i \, \omega \, \mu(\omega) \textbf{H}_S + \nabla \times \textbf{E}_S & = & i \omega (\mu(\omega) - \mu_b) \textbf{H}_{\mbox{inc}}, \medskip \\ 
+ \mbox{ Sommerfeld condition},
\end{array}
\right. 
\end{equation*}
where $\textbf{E}_{\mbox{inc}}, \textbf{H}_{\mbox{inc}}$ is the incident field, and $\varepsilon_b, \mu_b$ the background indices. The incident fields $\textbf{E}_{\mbox{inc}}, \textbf{H}_{\mbox{inc}}$ satisfy homogeneous Maxwell's equations with indices $\varepsilon_b, \mu_b$. Let us introduce 
$$ \textbf{J} = i \omega (\varepsilon(\omega) - \varepsilon_b) \, \textbf{E}_{\mbox{inc}} $$
and we consider only dielectric media such that $\mu(\omega) = \mu_b = \mu_0$ in the physical domain. As a result, the Maxwell's equations that will considered in the sequel are given as
\begin{equation}
\left \{ 
\begin{array}{lll}
-i \, \omega \, \varepsilon(\omega) \textbf{E}_S - \nabla \times \textbf{H}_S & = & \textbf{J}, \medskip \\
-i \, \omega \, \mu(\omega) \textbf{H}_S + \nabla \times \textbf{E}_S & = & 0, \medskip \\ 
+ \mbox{ Sommerfeld condition }.
\end{array}
\right. 
\label{eq:MaxwellSystem}
\end{equation}
Efficiently computing this scattered field for a large number of pulsations consists in expanding the solution into the QNM basis :
\begin{equation*}
\left[ \textbf{E}_S(\textbf{r},\omega), \textbf{H}_S(\textbf{r},\omega)\right] = \sum_m \alpha_m
(\omega) \left[ \widetilde{\textbf{E}}_m(\textbf{r},\omega),
 \widetilde{\textbf{H}}_m(\textbf{r}
,\omega)\right],
\end{equation*}
where the $\alpha_m$'s are the complex modal excitation coefficients, which measure how much the QNMs are excited by the driving field illuminating the resonator with a real frequency $\omega$. Note that we use a tilde to differentiate the QNM fields from other fields, for instance the scattered or driving fields, and consistently, we will also use a tilde to denote the QNM frequency $\tilde{\omega}_m$, in contrast to the real excitation frequency $\omega$. 

There are still some complicated mathematical issues in relation with the actual physical problem for which the open space is infinite and Maxwell’s operator are continuous. For instance, the conditions under which the completeness of the QNM expansions of Eq. (1) is guaranteed are not still fully understood \cite{bonod,gralak}. There are also several known and correct expressions for the $\alpha_m$'s \cite{LalanneReview}, but we do not know which offer the best performance, e.g. the fastest convergence rate towards the actual solution as the number of QNMs retained in the expansion increases. 

However, for practical geometries of interest in nanophotonics, the QNMs are computed numerically and it would be unrealistic to expect computing many QNMs over a broad spectral range, ideally in the entire lower half-plane of the complex plane $(Im(\tilde{\omega}_m)<0)$. Rather we have to consider a discretized version of the initial Maxwell's equations and the physical domain is bounded by perfectly-matched layers (PMLs). The discretized operator is a matrix of finite dimension, and its spectrum is composed of a finite number of QNMs (often the relevant ones involved in the resonator dynamics in the spectral range of interest) completed by a large number of PML modes, which have much less physical significance but warrant completeness \cite{Vial,Wei,LalanneReview}.

Efficient QNMs solvers exist for computing and normalizing QNMs and PML modes for various geometries, such as plasmonic crystals, metal gratings and plasmonic nanoantennas \cite{Gras19}; even freeware \cite{Bai13} or improved commercial software packages \cite{Wei} can be used. Thus the important remaining step is the reconstruction problem, i.e. the computation of the modal coefficients $\alpha_m$'s and the reconstruction of the scattered field. In this paper, we focus on material systems whose relative permittivity $\varepsilon(\omega)$ is described by a N-pole Lorentz permittivity (see \cite{LorentzOptic}):
\begin{equation}
\varepsilon(\omega)/\varepsilon_{\infty}=1-\sum_{i=1}^N\omega^2_{p,i}/(\omega^2-\omega^2_{0,i}+i\omega\gamma_i),
\label{eq:DrudeLorentz}
\end{equation}
which may model a large variety of systems with increasing accuracy as the number of poles increases. This model permits the introduction of auxiliary fields in order to linearize the previously-non-linear eigenvalue problem. It also respects the causality relation $\bar{\varepsilon}(\omega) = \varepsilon(-\bar{\omega})$ where $\bar{\omega}$ stands for the complex conjugate of $\omega$. The contribution of the free electron-gas of metals can be treated by a Drude permittivity, setting $\omega_{0,i}=0$.

Let us denote $\Omega_{res}$ the domain of the resonator for which $\varepsilon(\omega)$ is different from $\varepsilon_b$ (hence it is the support of the source term $\textbf{J}$).
In \cite{LalanneReview}, a review of the literature surrounding quasinormal modes, an attempt was made to classify the different formulas used to compute the excitation coefficients. At least, three different formulas for $\alpha_m$ were reported: 
\begin{itemize}
\item The formula 5.11 in \cite{LalanneReview}: 
\begin{equation}
\alpha_m = \dfrac{1}{i (\tilde{\omega}_m - \omega) } \int_{\Omega_{res}} \textbf{J}(\textbf{r}) \cdot \tilde{\textbf{E}}_m(\textbf{r})d \textbf{r} 
\label{eq:FormulaAlpha}
\end{equation}
\item The formula proposed in \cite{Wei} (equivalent to formula 5.6 in \cite{LalanneReview}):
\begin{equation}
\alpha_m = \int_{\Omega_{res}} (\varepsilon_b - \varepsilon_\infty) \textbf{E}_{inc} \cdot \tilde{\textbf{E}}_m d\Omega + \dfrac{\tilde{\omega}_m}{\tilde{\omega}_m-\omega}\int_{\Omega_{res}} (\varepsilon(\tilde{\omega}_m) - \varepsilon_b) \textbf{E}_{inc} \cdot \tilde{\textbf{E}}_m d\Omega
\label{eq:FormuleWei}
\end{equation}
\item The formula proposed in \cite{Marseillais} (equivalent to formula 5.10 in \cite{LalanneReview}):
\begin{equation}
    \alpha_m = \dfrac{\omega}{i \, \tilde{\omega}_m (\tilde{\omega}_m - \omega) } \int_{\Omega_{res}} \textbf{J}(\textbf{r}) \cdot \tilde{\textbf{E}}_m(\textbf{r})d \textbf{r}
    \label{eq:FormuleMarseille}
\end{equation}
\end{itemize}
All these formulas hold if the modes $\tilde{\textbf{E}}_m$ are normalized as follows 
\begin{equation}
 \int_\Omega \dfrac{\partial (\tilde{\omega}_m \, \varepsilon(\tilde{\omega}_m))}{\partial \tilde{\omega}_m} \tilde{\textbf{E}}_m \cdot \tilde{\textbf{E}}_m - \dfrac{ \partial \left(\tilde{\omega_m} \mu(\tilde{\omega}_m) \right)}{\partial \tilde{\omega_m}} \, \tilde{\textbf{H}}_m \cdot \tilde{\textbf{H}}_m d\Omega = 1.
\label{eq:Norm}
\end{equation}
where $\Omega$ is the computational domain. This is the usual normalization \cite{Muljarov18, Sauvan13, Bai13}.

\subsection{Discrete modal expansion}

In this paper, we propose a common formalism based on the discrete Maxwell's equations to obtain these three formulas that we show to be valid for both QNMs and PML modes. 
 More precisely, when $\varepsilon(\omega)$ is a rational function, auxiliary unknowns can be introduced in order to obtain a linear eigenvalue problem.
 After this linearization procedure and after discretization (e.g. with Finite Element Method), the time-harmonic Maxwell's Equations can be written
\begin{equation}
-i\omega \textbf{M}_h \textbf{U}_h + \textbf{K}_h \textbf{U}_h = \textbf{F}_h ,
\label{eq:FormulaFEM}
\end{equation}
where $\textbf{M}_h$ is the mass matrix, $\textbf{K}_h$ is the stiffness matrix, and $\textbf{F}_h$ is the source term (h denotes the mesh size).
$\textbf{U}_h$ is the main unknown that will contain components of $\textbf{E}$ and other auxiliary unknowns introduced to obtain a linear eigenvalue problem. 
The matrices $\textbf{M}_h$ and $\textbf{K}_h$ are independent of $\omega$, an example of matrices will be given in section \ref{sec:Core}.
 From a discrete point of view, once the discrete linear system \eqref{eq:FormulaFEM} is set, the biorthogonal projection of the unknown $\textbf{U}_h$ provides an unique formula for $\alpha_m$:
\begin{equation}
    \alpha_m = \dfrac{1}{i(\tilde{\omega}_m-\omega)} \langle \textbf{F}_h,\textbf{x}^\bot_m\rangle,
    \label{eq:AlphaDiscrete}
\end{equation} 
where $\textbf{x}^\bot_m$ is the left eigenvector (i.e. the conjugate of the biorthogonal). This biorthogonal projection is obtained by considering the relation \eqref{eq:FormulaFEM} and taking the scalar product with the left eigenvector. Details are given in section \ref{sec:Core}.
$\textbf{x}^\bot_m$ solves the transpose eigenvalue problem
$$ \textbf{K}_h^T \textbf{x}^\bot_m = i \tilde{\omega}_m \textbf{M}_h^T \textbf{x}^\bot_m. $$
In this paper, the convention $\langle x,y\rangle = \sum x_i y_i $ is used.  The formula \eqref{eq:AlphaDiscrete} holds if the eigenvectors $\textbf{x}_m$ are normalized such that
\begin{equation}
    \langle \textbf{M}_h \textbf{x}_m, \textbf{x}^\bot_m \rangle = 1,
    \label{eq:NormDiscrete}
\end{equation}
which is the discrete equivalent of \eqref{eq:Norm}. This result is proven in section \ref{sec:Core}. In that section, the proposed matrices $\textbf{M}_h$ and $\textbf{K}_h$ are symmetric, such that we have
$$ \textbf{x}_m^\bot = \textbf{x}_m .$$
 An infinity of formulas can be found by writing different linearizations of Maxwell's equations. Each different linearization will produce a new set of auxiliary unknowns, and consequently a different set of matrices $\textbf{K}_h$ and $\textbf{M}_h$ and right hand side $\textbf{F}_h$. The three aforementioned formulas are obtained as follows:
\begin{itemize}
\item The formula \eqref{eq:FormulaAlpha} is obtained by a direct linearization of system \eqref{eq:MaxwellSystem}. This derivation is detailed in section \ref{sec:Core}.
\item The formula \eqref{eq:FormuleWei} is obtained by choosing a different source $\textbf{F}_h$. This is the object of section \ref{sec:ComparWei}.
\item The formula \eqref{eq:FormuleMarseille} is obtained by starting from the second-order formulation of Maxwell's equations with curl-curl operator. This derivation is detailed in section \ref{sec:ComparMarseille}.
\end{itemize}
Other formulas exist \cite{LalanneReview} but will not be analyzed here. More recently, a newly developed formula is presented in \cite{Tong19}. An infinite set of formulas can be found by splitting the source on the different fields. For the linearization given in section \ref{sec:Core}, by writing the generalized source term as $\textbf{F} = [\textbf{f}_1, \textbf{f}_2, \textbf{f}_3, \textbf{f}_4]^T$, we can find the following generalization of the modal excitation coefficient: 
\begin{equation}
\alpha_m = \dfrac{1}{i(\omega_m - \omega)}\int_{\Omega_{Res}} \textbf{f}_1\cdot \tilde{\textbf{E}}_m + \textbf{f}_2 \cdot \tilde{\textbf{H}}_m + (\varepsilon(\tilde{\omega}_m)-\varepsilon_\infty)(\textbf{f}_3-i\tilde{\omega}_m\textbf{f}_4)\cdot \tilde{\textbf{E}}_m d\Omega 
\label{eq:GeneralFormula}
\end{equation}
provided that 
$$
-i \omega \textbf{f}_1 + i\omega (\varepsilon(\omega)-\varepsilon_\infty)(i\omega \textbf{f}_4 - \textbf{f}_3) - \nabla \times \left( \dfrac{1}{\mu_0} \textbf{f}_2 \right) = -i \omega \textbf{J} . 
$$
The derivation is detailed in section \ref{sec:SplitSource}. The modal solution is given as
\begin{equation}
\textbf{E}_{S}^{\mbox{modal}} = \sum_{m=1}^N \alpha_m \tilde{\textbf{E}}_m
\label{eq:modal_expansion}
\end{equation}
where $N$ is the number of modes conserved. The four formulas \eqref{eq:FormulaAlpha}, \eqref{eq:FormuleMarseille}, \eqref{eq:FormuleWei} and \eqref{eq:GeneralFormula} for coefficients $\alpha_m$ will  provide a field $\textbf{E}_{S}^{\mbox{modal}}$ that will converge to the scattered field $\textbf{E}_S$ when $N$ tends to the size of matrix $\textbf{M}_h$. Their convergence rate, however, may differ. 

In section \ref{sec:Degenerate}, it is explained how degenerate eigenvalues (i.e. multiple eigenvalues) can be treated correctly with a simple Gram-Schmidt orthogonalization procedure with respects to matrix $\textbf{M}_h$. In most of papers in the literature, eigenvalues are assumed to be simple. However, as the numerical results presented in \ref{sec:Numeric} show, there can be a non negligible number of degenerate eigenvalues.

The computational domain has to be truncated, e.g. with Perfectly Matched Layers. In order to keep real matrices $\textbf{M}_h$ and $\textbf{K}_h$ (and complex conjugate eigenvalues), dispersive PMLs have been chosen. The indexes $\varepsilon(\omega), \mu(\omega)$ are rational functions of $\omega$, they are given by formula \eqref{eq:EpsMuPML} in 3-D. In section \ref{sec:PML}, we detail how Maxwell's equations are linearized with respect to $\omega$, leading to non-symmetric matrices $\textbf{M}_h$ and $\textbf{K}_h$. Because the final eigenvalue problem solved by $\tilde{\textbf{E}}_m$ is symmetric, the left eigenvector $\textbf{x}^\bot_m$ can be computed directly from the right eigenvector $\textbf{x}_m$, formulas are given in section \ref{sec:PML}. The normalization \eqref{eq:Norm} is also valid for dispersive PMLs. The computational domain $\Omega$ involved in the integral includes both the physical domain and the PMLs.

Finally, numerical results are presented in section \ref{sec:Numeric} in order to compare the accuracy of the three formulas \eqref{eq:FormulaAlpha}, \eqref{eq:FormuleWei} and \eqref{eq:FormuleMarseille}.

\section{Eigenmode expansion for first-order formulation of Maxwell's equations}

\label{sec:Core}
In this section, we note $\textbf{E}, \textbf{H}$ the solutions of Maxwell's system \eqref{eq:MaxwellSystem}.

\subsection{Discrete expansion}

For the sake of illustration, we consider an isotropic (to simplify) medium with a dispersive permittivity described by the single-pole Lorentz model, 
$$\varepsilon(\omega)=\varepsilon_\infty \left(1-\dfrac{\omega_p^2}{\omega^2-\omega_0^2+i\gamma\omega} \right)$$ 
and a nondispersive permeability $\mu(\omega)=\mu_0$. We introduce two auxiliary fields, the polarization $\textbf{P}=-\varepsilon_\infty \dfrac{\omega_p^2}{\omega^2-\omega_0^2+i\gamma\omega}\textbf{E}$ and $\textbf{Q}=-i\omega\textbf{P}$. With elementary algebraic manipulations, we can reformulate Maxwell's system \eqref{eq:MaxwellSystem} as the following source problem
\begin{equation}
\left \{ 
\begin{array}{lll}
-i \, \omega \, \varepsilon_\infty \, \textbf{E}  + \textbf{Q} - \nabla \times \textbf{H} & = & \textbf{J} \medskip \\
-i \, \omega \, \mu_0 \, \textbf{H} + \nabla \times \textbf{E} & = & 0 \medskip \\ 
-i\omega\textbf{P} - \textbf{Q} &= & 0 \medskip \\
i\omega\textbf{Q}-\gamma\textbf{Q}-\omega_0^2\textbf{P} +\varepsilon_\infty \omega_p^2 \textbf{E}& = & 0  \medskip \\
+ \mbox{ Sommerfeld condition}
\end{array}
\right. 
   \label{eq:MaxwellSystemPQ}
\end{equation}
In order to obtain a symmetric system, we multiply the second equation by $-1$, the third equation by $\omega_0^2 /(\varepsilon_\infty \omega_p^2)$ and the fourth by $1/(\varepsilon_\infty \omega_p^2)$. 
$$
\left \{ 
\begin{array}{lll}
-i \, \omega \, \varepsilon_\infty \, \textbf{E}  + \textbf{Q} - \nabla \times \textbf{H} & = & \textbf{J} \medskip \\
+i \, \omega \, \mu_0 \, \textbf{H} - \nabla \times \textbf{E} & = & 0 \medskip \\ 
-i \omega \dfrac{\omega_0^2}{\varepsilon_\infty\omega^2_p}\textbf{P}-\dfrac{\omega_0^2}{\varepsilon_\infty\omega^2_p}\textbf{Q} &= & 0 \medskip \\
\dfrac{i \omega}{\varepsilon_\infty \omega^2_p} \textbf{Q} - \dfrac{\gamma}{\varepsilon_\infty \omega_p^2}\textbf{Q} - \dfrac{\omega_0^2}{\varepsilon_\infty \omega_p^2}\textbf{P} + \textbf{E} & = & 0  \medskip \\
+ \mbox{ Sommerfeld condition}
\end{array}
\right. 
$$
We can write this system using the linear operators $\textbf{K}$ and $\textbf{M}$
$$ \textbf{K} \textbf{U} - i\omega \textbf{M} \textbf{U} =  \textbf{F} $$
with 
$$
\textbf{K}= \left[
\begin{array}{cccc}
     0 & -\nabla\times & 0 & 1 \\
     -\nabla\times & 0  & 0 & 0 \\
     0 & 0 & 0 & -\dfrac{\omega_0^2}{\varepsilon_\infty \omega_p^2} \\
     1 & 0 & -\dfrac{\omega_0^2}{\varepsilon_\infty \omega_p^2} & - \dfrac{\gamma}{\varepsilon_\infty \omega_p^2} 
\end{array}
\right],
$$
$$
\textbf{M}= \left[
\begin{array}{cccc}
     \varepsilon_\infty & 0 & 0 & 0 \\
     0 & -\mu_0  & 0 & 0 \\
     0 & 0 & \dfrac{\omega_0^2}{\varepsilon_\infty\omega_p^2} & 0 \\
     0 & 0 & 0 & - \dfrac{1}{\varepsilon_\infty\omega_p^2} 
\end{array}
\right],
\quad \textbf{F} = \left[\begin{array}{l} \textbf{J} \\ 0 \\ 0 \\ 0 \end{array} \right]
$$
~\\
After discretization, the Maxwell's system is then given as
 \begin{equation}
-i \omega \textbf{M}_h \textbf{U}_h + \textbf{K}_h \textbf{U}_h = \textbf{F}_h
\label{eq:DiscreteMaxwell}
 \end{equation} 
 where 
 $ \textbf{U}_h = \left( \textbf{E}_h, \textbf{H}_h, \textbf{P}_h, \textbf{Q}_h \right), $
 and $\textbf{E}_h, \textbf{H}_h, \textbf{P}_h, \textbf{Q}_h$ contain the components of $\textbf{E}, \textbf{H}, \textbf{P}, \textbf{Q}$ on basis functions. The source term $\textbf{F}_h$ is given as
 $$ (\textbf{F}_h)_i = \int_{\Omega_{res}} \textbf{J}(\textbf{r}) \cdot \boldsymbol{\varphi}_i(\textbf{r}) \, d\textbf{r} $$
 where $\boldsymbol{\varphi}_i$ are basis functions for unknown $\textbf{E}_h$. Matrices $\textbf{M}_h$ and $\textbf{K}_h$ are given in appendix \ref{app:FemMatrices}.
 The right eigenvectors $\textbf{x}_m$ solve the eigenproblem
 \begin{equation}
 \textbf{K}_h \textbf{x}_m = \lambda_m \textbf{M}_h \textbf{x}_m 
 \label{eq:DiscreteEigen}
  \end{equation} 
 where the eigenvalue $\lambda_m$ is linked with $\tilde{\omega}_m$ by
 $$ \lambda_m = i \tilde{\omega}_m $$
 Assuming that $\textbf{M}_h^{-1} \textbf{K}_h$ is diagonalizable, we have
 $$ \textbf{M}_h^{-1} \textbf{K}_h = \textbf{V} \textbf{D} \textbf{V}^{-1} $$
 where $\textbf{D}$ is a diagonal matrix with eigenvalues $\lambda_m$ on the diagonal and $\textbf{V}$ the matrix whose columns are formed with right eigenvectors $\textbf{x}_m$. The left eigenvectors of $\textbf{M}_h^{-1} \textbf{K}_h$ denoted $\textbf{w}_m$ are the rows of matrix $\textbf{V}^{-1}$. Since $\textbf{V} \textbf{V}^{-1} = \textbf{I}$, vectors $\textbf{x}_m$ and $\textbf{w}_m$ are biorthogonal
 $$ \langle \textbf{x}_m , \textbf{w}_n \rangle = \delta_{m, n} $$
 The left eigenvectors $\textbf{w}_m$ can also be found by searching right eigenvectors of the transpose of $\textbf{M}_h^{-1} \textbf{K}_h$. Since $\textbf{K}_h$ and $\textbf{M}_h$ are symmetric, we have
 $$ (\textbf{M}_h^{-1} \textbf{K}_h)^T = \textbf{K}_h \textbf{M}_h^{-1} $$
 Hence $\textbf{w}_m$ solves the following eigenvalue problem
 $$ \textbf{K}_h \textbf{M}_h^{-1} \textbf{w}_m = \lambda_m \textbf{w}_m $$
 By introducing $\textbf{x}^\bot_m = \textbf{M}_h^{-1} \textbf{w}_m$, we obtain
 $$ \textbf{K}_h \textbf{x}^\bot_m = \lambda_m \textbf{M}_h \textbf{x}^\bot_m $$ 
 $\textbf{x}^\bot_m$ is the left eigenvector of generalized eigenproblem \eqref{eq:DiscreteEigen}.
 If $\lambda_m$ is a simple eigenvalue, $\textbf{x}^\bot_m$ is colinear with $\textbf{x}_m$ since they solve the same eigenvalue problem. In order to have $\textbf{x}^\bot_m = \textbf{x}_m$, the eigenvector $\textbf{x}_m$ must be normalized such that
 \begin{equation}
 \langle \textbf{M}_h \textbf{x}_m, \textbf{x}_m \rangle = 1
 \end{equation}
 The solution $\textbf{U}_h$ is expanded with right eigenvectors $\textbf{x}_m$ (they form a basis since the matrix is diagonalizable):
 $$ \textbf{U}_h = \sum_m \alpha_m \textbf{x}_m $$
 By injecting this expansion in \eqref{eq:DiscreteMaxwell} and using \eqref{eq:DiscreteEigen}), we obtain
 $$ \sum_m \alpha_m  (- i \, \omega + i \, \tilde{\omega}_m)  \textbf{M}_h \textbf{x}_m =  \textbf{F}_h $$ 
 The modal coefficient $\alpha_m$ is directly obtained by taking the scalar product $\langle, \rangle$ with the left eigenvector $\textbf{x}^\bot_m$
 $$ \alpha_m (-i \, \omega + i \, \tilde{\omega}_m) = \langle \textbf{F}_h, \textbf{x}^\bot_m \rangle $$
 Since $\textbf{x}^\bot_m = \textbf{x}_m$, we obtain
 \begin{equation}
     \alpha_m = \dfrac{1}{i(\tilde{\omega}_m - \omega)} \langle \textbf{F}_h, \textbf{x}_m \rangle
     \label{eq:AlphaDiscrete2}
 \end{equation}  
 which is the announced result in the introductio, implying that the expansion coefficient solely depends on the QNM and not on the left eigenvector. This important results provides analyticity which has not been obtained in the related works by \cite{Vial} and was derived in a different way using the divergence theorem and the continuous operator, not the discretized one, in \cite{Wei}.

\subsection{Link with continuous expansion}

The formula \eqref{eq:AlphaDiscrete2} is the discrete equivalent of \eqref{eq:FormulaAlpha}
since 
$$ \langle \textbf{F}_h, \textbf{x}_m \rangle = \sum_i x_{m, i} \int_{\Omega_{res}} \textbf{J}(\textbf{r}) \cdot \boldsymbol{\varphi}_i(\textbf{r})  dr $$ 
where $x_{m, i}$ is the $i$-th component of $\textbf{x}_m$. By swapping the sum and the integral, we obtain
$$\langle \textbf{F}_h, \textbf{x}_m \rangle = \int_{\Omega_{res}} \textbf{J}(\textbf{r}) \cdot \textbf{x}_m(\textbf{r}) dr $$
For numerical experiments, it is preferable to perform a scalar product as presented in formula \eqref{eq:AlphaDiscrete2} rather than approximating this integral. With the same arguments, we have the following equality
$$ \langle \textbf{M}_h \textbf{x}_m , \textbf{x}_m \rangle = \int_{\Omega} \varepsilon_e \tilde{\textbf{E}}_m \cdot \tilde{\textbf{E}}_m - \mu_0 \tilde{\textbf{H}}_m \cdot \tilde{\textbf{H}}_m + \dfrac{\omega_0^2}{\varepsilon_\infty \omega_p^2} \tilde{\textbf{P}}_m \cdot \tilde{\textbf{P}}_m - \dfrac{1}{\varepsilon_\infty \omega_p^2} \tilde{\textbf{Q}}_m \cdot \tilde{\textbf{Q}}_m \, d\Omega $$
where
$$ \varepsilon_e =  \left \{ \begin{array}{l}
\varepsilon_\infty \, \mbox{ in } \Omega_{res} \\
\varepsilon_b, \mbox{ elsewhere. }
\end{array} \right. , 
\quad \textbf{x}_m = \left[ \begin{array}{c} \tilde{\textbf{E}}_m \\ \tilde{\textbf{H}}_m \\ \tilde{\textbf{P}}_m
 \\ \tilde{\textbf{Q}}_m \end{array} \right]. $$
Since $\tilde{\textbf{P}}_m = -\varepsilon_\infty  \omega_p^2 / (\tilde{\omega}_m^2 + i \gamma \tilde{\omega}_m - \omega_0^2) \tilde{\textbf{E}}_m$ and $\tilde{\textbf{Q}}_m = -i \tilde{\omega}_m \tilde{\textbf{P}}_m$, we get
$$ \dfrac{\omega_0^2}{\varepsilon_\infty \, \omega_p^2} \tilde{\textbf{P}}_m \cdot \tilde{\textbf{P}}_m - \dfrac{1}{\varepsilon_\infty \omega_p^2} \tilde{\textbf{Q}}_m \cdot \tilde{\textbf{Q}}_m = \varepsilon_\infty  \omega_p^2 
\dfrac{\left(\tilde{\omega}_m^2 + \omega_0^2 \right) }{\left( \tilde{\omega}_m^2 + i \gamma \tilde{\omega}_m - \omega_0^2\right)^2}  \tilde{\textbf{E}}_m \cdot \tilde{\textbf{E}}_m $$
Since we have
$$ \dfrac{\partial \varepsilon(\omega)}{\partial \omega} = \dfrac{\omega_p^2 \varepsilon_\infty (2 \omega + i \gamma)}{\left(\omega^2 - \omega_0^2 + i \gamma \omega \right)^2} $$
we obtain
$$ \dfrac{\partial \left( \tilde{\omega}_m \varepsilon(\tilde{\omega}_m) \right)}{ \partial \tilde{\omega}_m} = \left \{ \begin{array}{l}
\varepsilon_\infty +  \varepsilon_\infty \omega_p^2 \dfrac{\tilde{\omega}_m^2 + \omega_0^2}{\left( \tilde{\omega}_m^2 + i \gamma \tilde{\omega}_m - \omega_0^2 \right)^2 }, \mbox{ in } \Omega_{res} \medskip \\ 
  \varepsilon_b, \mbox{ otherwise } \end{array} \right.
$$
As a result, we have proven that
$$ \langle \textbf{M}_h \textbf{x}_m, \textbf{x}_m \rangle = \int_{\Omega} \dfrac{\partial \left( \tilde{\omega}_m \varepsilon(\tilde{\omega}_m) \right) }{\partial \tilde{\omega}_m} \tilde{\textbf{E}}_m \cdot \tilde{\textbf{E}}_m - \mu_0 \tilde{\textbf{H}}_m \cdot \tilde{\textbf{H}}_m d \Omega. $$
This relation proves that the normalization \eqref{eq:NormDiscrete} is the discrete equivalent of \eqref{eq:Norm}.  Again, for the sake of simplicity, the relation \eqref{eq:NormDiscrete} is preferred to normalize discrete eigenvectors.
\begin{remark}
The normalization can be written with only unknown $\tilde{\textbf{E}}_m$.
By using the relation $\tilde{\textbf{H}}_m = \dfrac{1}{i \omega \mu_0} \nabla \times \tilde{\textbf{E}}_m$ and the variational formulation satisfied by $\tilde{\textbf{E}}_m$ with only Dirichlet or Neumann boundary conditions:
$$ -\tilde{\omega}_m^2 \int_\Omega \varepsilon(\tilde{\omega}_m) \tilde{\textbf{E}}_m \cdot \tilde{\textbf{E}}_m \, d \Omega 
+ \int_\Omega \dfrac{1}{\mu_0} \nabla \times \tilde{\textbf{E}}_m \cdot \nabla \times \tilde{\textbf{E}}_m \, d\Omega = 0, $$
we obtain that
$$ - \int_\Omega \mu_0 \, \tilde{\textbf{H}}_m \cdot \tilde{\textbf{H}}_m \, d\Omega = \int_\Omega \varepsilon(\tilde{\omega}_m) \tilde{\textbf{E}}_m \cdot \tilde{\textbf{E}}_m \, d\Omega. $$
As a result the normalization can be written as
$$ \langle \textbf{M}_h \textbf{x}_m, \textbf{x}_m \rangle = \int_\Omega \dfrac{\partial \left( \tilde{\omega}_m \varepsilon(\tilde{\omega}_m) \right) }{\partial \tilde{\omega}_m} \tilde{\textbf{E}}_m \cdot \tilde{\textbf{E}}_m
 + \varepsilon(\tilde{\omega}_m) \tilde{\textbf{E}}_m \cdot \tilde{\textbf{E}}_m d\Omega. $$
\end{remark}

\section{Derivation of other formulas and issues}

\subsection{Derivation of formula of \cite{Wei}}
\label{sec:ComparWei}

To obtain the formula \eqref{eq:FormulaAlpha}, first, we have written Maxwell's equations directly for the scattered field $\textbf{E}_S(\textbf{r},\omega), \textbf{H}_S(\textbf{r},\omega)$ and then introduced the auxiliary fields $\textbf{P}$ and $\textbf{Q}$. In the aforementioned paper \cite{Wei},
 Maxwell's equations are first written for the total field, and the auxiliary unknowns $\textbf{P}$ and $\textbf{Q}$ are introduced at this step. Hence the unknowns $\textbf{E}, \textbf{H}, \textbf{P}, \textbf{Q}$ solve the system \eqref{eq:MaxwellSystemPQ} with $\textbf{J} = 0$. As a second step, we subtract the equations solved by the incident field (homogeneous Maxwell's equation with indices $\varepsilon_b$ and $\mu_0$),  and use the relations
 $$ \left[ \textbf{E}(\textbf{r},\omega), \textbf{H}(\textbf{r},\omega)\right] = \left[ \textbf{E}_S(\textbf{r},\omega) + \textbf{E}_{\mbox{inc}}(\textbf{r},\omega), \; \textbf{H}_S(\textbf{r},\omega) + \textbf{H}_{\mbox{inc}}(\textbf{r},\omega)\right]$$
 to obtain the system solved by the scattered field
\begin{equation}
\left \{ 
\begin{array}{lll}
-i \, \omega \, \varepsilon_\infty \, \textbf{E}_S  + \textbf{Q}_S - \nabla \times \textbf{H}_S & = & i\omega(\varepsilon_\infty - \varepsilon_b)\textbf{E}_{\mbox{inc}} \medskip \\
+i \, \omega \, \mu_0 \, \textbf{H}_S - \nabla \times \textbf{E}_S & = & 0 \medskip \\ 
-i \omega \dfrac{\omega_0^2}{\varepsilon_\infty\omega^2_p}\textbf{P}_S-\dfrac{\omega_0^2}{\varepsilon_\infty\omega^2_p}\textbf{Q}_S &= & 0 \medskip \\
\dfrac{i \omega}{\varepsilon_\infty \omega^2_p} \textbf{Q}_S - \dfrac{\gamma}{\varepsilon_\infty \omega_p^2}\textbf{Q}_S - \dfrac{\omega_0^2}{\varepsilon_\infty \omega_p^2}\textbf{P}_S + \textbf{E}_S & = & -\textbf{E}_{\mbox{inc}}  \medskip \\
+ \mbox{ Sommerfeld condition}
\end{array}
\right. 
\label{eq:SystemWei}
\end{equation}

Unlike the equations considered in section \ref{sec:Core}, we can see that the source term on the right hand side of the equations is no longer confined to the first equation.  
%
%
The coefficient $\alpha_m$ becomes:
%
$$\alpha_m = \int_{\Omega_{res}} (\varepsilon_b - \varepsilon_\infty) \textbf{E}_{\mbox{inc}} \cdot \tilde{\textbf{E}}_m d\Omega + \dfrac{\tilde{\omega}_m}{\tilde{\omega}_m-\omega}\int_{\Omega_{res}} (\varepsilon(\tilde{\omega}_m) - \varepsilon_b) \textbf{E}_{\mbox{inc}} \cdot \tilde{\textbf{E}}_m d\Omega .$$
It is important to notice that the systems \eqref{eq:SystemWei} and \eqref{eq:MaxwellSystemPQ} provide exactly the same numerical solution $\textbf{E}_S$. Only the auxiliary fields $\textbf{P}$ and $\textbf{Q}$ differ, that's why the source $\textbf{F}_h$ is different between the two approaches and two different formulas are obtained for $\alpha_m$. Other formulas for $\alpha_m$ can be found by choosing a different distribution of the source over the four equations. This is the object of the next sub-section.

\subsection{Generalized Sources}
\label{sec:SplitSource}

Let us split the source term $\textbf{J}$ into a set of artificial sources denoted $\textbf{f}_1, \textbf{f}_2, \textbf{f}_3, \textbf{f}_4$.  
$$
\left \{ 
\begin{array}{lll}
-i \, \omega \, \varepsilon_\infty \, \textbf{E}  + \textbf{Q} - \nabla \times \textbf{H} & = & \textbf{f}_1 \medskip \\
+i \, \omega \, \mu_0 \, \textbf{H} - \nabla \times \textbf{E} & = & \textbf{f}_2 \medskip \\ 
-i \omega \dfrac{\omega_0^2}{\varepsilon_\infty\omega^2_p}\textbf{P}-\dfrac{\omega_0^2}{\varepsilon_\infty\omega^2_p}\textbf{Q} &= & \textbf{f}_3 \medskip \\
\dfrac{i \omega}{\varepsilon_\infty \omega^2_p} \textbf{Q} - \dfrac{\gamma}{\varepsilon_\infty \omega_p^2}\textbf{Q} - \dfrac{\omega_0^2}{\varepsilon_\infty \omega_p^2}\textbf{P} + \textbf{E} & = & \textbf{f}_4  \medskip \\
+ \mbox{ Boundary conditions}
\end{array}
\right. 
$$
By eliminating the unknowns \textbf{H}, \textbf{P}, and \textbf{Q}, we obtain the following equation for \textbf{E}: 
$$
-\omega^2 \varepsilon(\omega) \textbf{E} + \nabla \times \left( \dfrac{1}{\mu_0} \nabla \times \textbf{E} \right) = -i\omega \textbf{f}_1 + \dfrac{i\omega \varepsilon_\infty \omega_p^2}{-\omega^2-i\omega\gamma+\omega_0^2}(i\omega \textbf{f}_4 - \textbf{f}_3) - \nabla \times \left( \dfrac{1}{\mu_0} \textbf{f}_2 \right)
$$
which is equivalent to the standard Maxwell's equations: 
$$
\omega^2 \varepsilon(\omega) \textbf{E} + \nabla \times \left( \dfrac{1}{\mu_0} \nabla \times \textbf{E} \right) = -i\omega \textbf{J}
$$
as soon as 
$$
-i \omega \textbf{f}_1 + \dfrac{i\omega \varepsilon_\infty \omega_p^2}{-\omega^2-i\omega\gamma+\omega_0^2}(i\omega \textbf{f}_4 - \textbf{f}_3) - \nabla \times \left( \dfrac{1}{\mu_0} \textbf{f}_2 \right) = -i \omega \textbf{J}.
$$
By choosing different splittings of the source (i.e. different functions $\textbf{f}_1, \textbf{f}_2, \textbf{f}_3, \textbf{f}_4$ that satisfy the relationship above), we will obtain different formulas for $\alpha_m$. The modal solution obtained with these different formulas (see equation \eqref{eq:modal_expansion}) will converge towards the same electric field $\textbf{E}_S$ when the number of modes is increased.

\subsection{Derivation of formula in \cite{Marseillais}}
\label{sec:ComparMarseille}

In this section we propose a different linearization of the problem by starting from the second order formulation. With this alternative linearization, we obtain the formula \eqref{eq:FormuleMarseille} for the coefficients $\alpha_m$. Let us start from the second-order formulation of Maxwell's equations

$$-\omega^2 \varepsilon(\omega) \textbf{E} + \nabla\times \left(\dfrac{1}{\mu_0}\nabla\times \textbf{E}\right)=-i\omega \textbf{J} .$$
In order to linearize this equation, let us introduce the field $\textbf{E}' = -i \omega \textbf{E}$ and the auxiliary field $\textbf{P} = \left( \varepsilon(\omega)-\varepsilon_\infty \right) \textbf{E}'$ and $\textbf{Q} = -i \omega \textbf{P}$. We obtain the following system of linear equations: 
$$
\left\{
\begin{array}{lll}
-i \, \omega \textbf{E} -\textbf{E}'     & =  & 0 \\
-i \, \omega \varepsilon_\infty \textbf{E}' + \textbf{Q} + \nabla \times \left(\dfrac{1}{\mu_0}\nabla\times \textbf{E}\right)     & = & - i \, \omega \textbf{J}  \\
 -i \, \omega \textbf{P} - \textbf{Q}    & =  & 0  \medskip \\
    -i \omega \textbf{Q} + \gamma \textbf{Q}  + \omega_0^2 \textbf{P} - \varepsilon_\infty 
    \omega_p^2 \textbf{E}' & =  & 0
    
\end{array}
\right. ,
$$
which gives the following stiffness and mass operators $\textbf{K}$ and $\textbf{M}$ for the vector $\textbf{U} = [\textbf{E}, \textbf{E}', \textbf{P}, \textbf{Q}]^T$:
$$
\textbf{K}= \left[
\begin{array}{cccc}
     0 & -1 & 0 & 0 \\
     \dfrac{1}{\mu_0} \nabla\times \nabla\times & 0  & 0 & 1 \\
     0 & 0 & 0 & -1 \\
     0 & -\varepsilon_\infty \omega_p^2 & \omega_0^2 & \gamma 
\end{array}
\right],
$$
$$
\textbf{M} = \left[
\begin{array}{cccc}
    1 & 0 & 0 & 0 \\
    0 & \varepsilon_\infty & 0 & 0 \\
    0 & 0 & 1 & 0 \\
    0 & 0 & 0 & 1 
\end{array}{}
\right] .
$$
As a result, Maxwell's equations are rewritten as :
$$
(-i\, \omega \textbf{M} + \textbf{K}) \textbf{U} = \textbf{F} ,
$$
 where
 $$ \textbf{F} = [ 0, -i \omega \textbf{J}, 0, 0] $$
 is the source term. After discretization, we have the following discrete system
$$
 (-i \omega \textbf{M}_h + \textbf{K}_h)  \textbf{U}_h = \textbf{F}_h .$$
The matrices $\textbf{M}_h, \textbf{K}_h$ are not detailed here, but are different from matrices 
 $\textbf{M}_h$ and $\textbf{K}_h$ given in section \ref{sec:Core}.
 It can be noticed that the discrete solution $\textbf{E}_h$ will be exactly the same 
 with this formulation or with the formulation presented in section \ref{sec:Core}.
 The right eigenvectors $\textbf{x}_m$ solve the eigenvalue problem
 $$\textbf{K}_h \textbf{x}_m = i \tilde{\omega}_m \textbf{M}_h \textbf{x}_m $$
 while the left eigenvectors $\textbf{x}^\bot_m$ solve the adjoint eigenvalue problem
 $$ \textbf{K}_h^T \textbf{x}^\bot_m = i \tilde{\omega}_m \textbf{M}_h^T \textbf{x}^\bot_m. $$
Since we have 
$$
\textbf{K}^T= \left[
\begin{array}{cccc}
     0 & \dfrac{1}{\mu_0} \nabla\times \nabla\times & 0 & 0 \\
     -1 & 0  & 0 & -\varepsilon_\infty \omega_p^2 \\
     0 & 0 & 0 & \omega_0^2 \\
     0 & 1 & -1 & \gamma 
\end{array}
\right],
$$
$$
\textbf{M}^T = \textbf{M}
$$
we obtain the following system of equations for the biorthogonal eigenvectors ($\textbf{x}^\bot_m = [\textbf{E}_\bot, \textbf{E}'_\bot, \textbf{P}_\bot, \textbf{Q}_\bot ]$): 
$$
\left\{
\begin{array}{lll}
-i \, \tilde{\omega}_m \textbf{E}_\bot +\nabla\times\left(\dfrac{1}{\mu_0}\nabla\times \textbf{E}'_\bot\right) & =  & 0  \medskip \\
  -i \, \tilde{\omega}_m \varepsilon_\infty \textbf{E}'_\bot -\textbf{E}_\bot - \varepsilon_\infty \omega_p^2 \textbf{Q}_\bot  & =  & 0 \medskip \\
    -i \, \tilde{\omega}_m \textbf{P}_\bot + \omega_0^2 \textbf{Q}_\bot & = & 0 \medskip \\
    -i \, \tilde{\omega}_m \textbf{Q}_\bot + \gamma \textbf{Q}_\bot + \textbf{E}'_\bot - \textbf{P}_\bot & = & 0, 
\end{array}
\right.
$$
By eliminating the other variables, we can show that $\textbf{E}'_\bot$ verifies 
$$
-\tilde{\omega}_m^2\varepsilon(\tilde{\omega}_m)\textbf{E}'_\bot + \nabla \times \left( \dfrac{1}{\mu_0} \nabla \times \textbf{E}'_\bot \right) = 0 ,
$$
and subsequentially : 
$$
\left\{
\begin{array}{lll}
\textbf{E}'_\bot     & = & \tilde{\textbf{E}}_m \medskip \\
\textbf{E}_\bot     & = &  -i\tilde{\omega}_m\varepsilon(\tilde{\omega}_m)\tilde{\textbf{E}}_m \medskip \\
\textbf{P}_\bot     & = & \dfrac{\omega_0^2}{\omega^2_0-i\gamma\tilde{\omega}_m-\tilde{\omega}_m^2}  \tilde{\textbf{E}}_m \medskip \\
\textbf{Q}_\bot & = & \dfrac{i\tilde{\omega}_m}{\omega^2_0-i\gamma\tilde{\omega}_m-\tilde{\omega}_m^2} \tilde{\textbf{E}}_m, \\
\end{array}
\right.
$$
where $\tilde{\textbf{E}}_m$ is the E-component of the the left eigenvector $\textbf{x}_m$.
We can now obtain the excitation coefficient :
$$
\alpha_m = \dfrac{1}{i \left(\tilde{\omega}_m - \omega \right) } \dfrac{\langle \textbf{F}_h, \textbf{x}^\bot_m \rangle}{\langle \textbf{M}_h \textbf{x}_m, \textbf{x}^\bot_m \rangle} = \dfrac{-i \omega \displaystyle \int_{\Omega_{res}} \textbf{J}(\textbf{r}) \cdot \tilde{\textbf{E}}_m(r) d \textbf{r} } {i(\tilde{\omega}_m-\omega) \, N_m},
$$
where the coefficient $N_m$ appears since we choose the normalization \eqref{eq:Norm} of the first order formulation. $N_m$ is given as
$$
N_m = \langle \textbf{M}_h \textbf{x}_m, \textbf{x}^\bot_m \rangle = \int_\Omega \tilde{\textbf{E}}_m \cdot \textbf{E}_\bot + \varepsilon_\infty \tilde{\textbf{E}}_m \cdot \textbf{E}'_\bot + \dfrac{\varepsilon_\infty \,  \omega_p^2}{\omega^2_0-i\gamma\tilde{\omega}_m-\tilde{\omega}_m^2}  \left( - i \omega_m \tilde{\textbf{E}}_m \cdot \textbf{P}_\bot - \tilde{\omega}_m^2 \tilde{\textbf{E}}_m\cdot \textbf{Q}_\bot \right) d \Omega.
$$
By substituting $\textbf{E}_\bot, \textbf{E}'_\bot, \textbf{P}_\bot, \textbf{Q}_\bot$ by the expressions above, we obtain
$$
N_m = - i \tilde{\omega}_m \left[ \int_\Omega \varepsilon(\tilde{\omega}_m) \tilde{\textbf{E}}_m \cdot \tilde{\textbf{E}}_m + \varepsilon_\infty \tilde{\textbf{E}}_m \cdot \tilde{\textbf{E}}_m + \dfrac{\varepsilon_\infty \omega_p^2}{(\omega^2_0-i\gamma\tilde{\omega}_m-\tilde{\omega}_m^2)^2} (\omega_0^2 \tilde{\textbf{E}}_m \cdot \tilde{\textbf{E}}_m + \tilde{\omega}_m^2 \tilde{\textbf{E}}_m\cdot \tilde{\textbf{E}}_m) \, d\Omega \right].
$$
We recognize the normalization used by the first order formulation multiplied by $-i \tilde{\omega}_m$. As a result, if $\tilde{\textbf{E}}_m$ is normalized by \eqref{eq:Norm}, we obtain that
$$ N_m = -i \tilde{\omega}_m, $$
which gives us this expression for the excitation coefficient: 

$$
\alpha_m = \dfrac{\omega}{i\tilde{\omega}_m(\tilde{\omega}_m-\omega)}\int_{\Omega_{res}} \textbf{J} \cdot \tilde{\textbf{E}}_m d\Omega.
$$.
We recognize the formula \eqref{eq:FormuleMarseille}.

\subsection{Treatment of degenerate eigenvalues}
\label{sec:Degenerate}

A set of degenerate modes $\{ \textbf{x}_k \}_{m_1\leq k \leq m_2}, $ are solutions of the eigenvalue problem at the same eigenfrequency $\tilde{\omega}_{m_1}$. Degenerate eigenvectors do not necessarily form an orthogonal sub-basis with respects to $\textbf{M}_h$. However, using Gram-Schmidt orthogonalization process, an orthogonal sub-basis with respects to $\textbf{M}_h$ can be constructed from the set of degenerate modes by algorithm \ref{algo:GramSchmidt}.
\begin{algorithm}
\caption{Algorithm to apply Gram-Schmidt orthogonalization to vectors $\textbf{x}_m$}
\label{algo:GramSchmidt}
\begin{algorithmic}
\FOR{m=$m_1$ to $m_2$}
\STATE{Initialize $\textbf{y} = \textbf{x}_m$}
\FOR{j = $m_1$ to $m-1$}
\STATE{Compute $\alpha = \langle \textbf{M}_h \textbf{x}_m, \textbf{x}^\bot_j \rangle$ }
\STATE{Substitute $\textbf{y}$ by $\textbf{y} - \alpha \textbf{x}_j$}
\ENDFOR   
\STATE{Compute left eigenvector $\textbf{y}^\bot$ from right eigenvector $\textbf{y}$ with formula \eqref{eq:LeftEigenVecPML3D} }
\STATE{Substitute $\textbf{x}_m$ by $\textbf{y} / \langle \textbf{M}_h \textbf{y}, \textbf{y}^\bot \rangle$}
\STATE{Store $\textbf{x}^\bot_j = \textbf{y}^\bot / \langle \textbf{M}_h \textbf{y}, \textbf{y}^\bot \rangle $}
\ENDFOR
\end{algorithmic}
\end{algorithm} 
%
By applying this procedure, the formula \eqref{eq:AlphaDiscrete} holds for degenerate eigenvalues with normalization \eqref{eq:NormDiscrete}. This process can also be done with continuous eigenmodes by replacing $\langle \textbf{M}_h \textbf{x}_m, \textbf{x}^\bot_j \rangle$ by 
$$
 \int_\Omega \dfrac{\partial (\tilde{\omega}_m \, \varepsilon(\tilde{\omega}_m))}{\partial \tilde{\omega}_m} \tilde{\textbf{E}}_m \cdot \tilde{\textbf{E}}_j - \dfrac{ \partial \left(\tilde{\omega_m} \mu(\tilde{\omega}_m) \right)}{\partial \tilde{\omega_m}} \, \tilde{\textbf{H}}_m \cdot \tilde{\textbf{H}}_j d\Omega .
$$
Here $\mu$ depends on $\omega$ inside the PML layers, which are detailed in the next sub-section.

\subsection{PML}
\label{sec:PML}

In this section, we describe how dispersive PMLs are handled.
The damping coefficients $\sigma_x$, $\sigma_y$ and $\sigma_z$ inside a PML where $x > x_0$, $y > y_0$ or $z>z_0$ are parabolic: 
$$
\sigma_1 = \sigma_x = \dfrac{3\, \text{log}(1000)}{2a^3} (x-x_0)^2 v_{max} \, \sigma
$$
$$
\sigma_2 = \sigma_y = \dfrac{3\, \text{log}(1000)}{2a^3} (y-y_0)^2 v_{max} \, \sigma
$$
$$
\sigma_3 = \sigma_z = \dfrac{3\, \text{log}(1000)}{2a^3} (z-z_0)^2 v_{max} \, \sigma.
$$
The coefficient $\sigma$ serves to adjust the reflection coefficient of the PML. $v_{max}$ is the speed of the wave inside the PML. 
In this section, we describe the formulation used for dispersive PMLs. The matrices $\textbf{M}_h, \textbf{K}_h$ are no longer symmetric.
 We provide relations between the left eigenvector $\textbf{x}_m^\bot$ and right eigenvector $\textbf{x}_m$. As a result we do not need to compute the eigenvectors of the adjoint problem, since we can compute $\textbf{x}^\bot_m$ directly from the right eigenvector $\textbf{x}_m$.

\subsubsection{2-D case}
\label{sec:PML2D}

In Transverse Electric case, we have
$$
\textbf{E} = u \, \textbf{e}_z, \quad \textbf{H} = v_x \, \textbf{e}_x + v_y \, \textbf{e}_y .$$
We use a split formulation of the PMLs where $u=u_1+u_2$ inside the PML. The unknowns $u_1$, $u_2$, and $\textbf{v}=(v_x, v_y)$ are solutions of:
$$
\left\{
\begin{array}{l}
     -i \omega \, \varepsilon_b \, u_1 + \varepsilon_b \, \sigma_x \, u_1 - \dfrac{\partial v_x}{\partial x}  =  0 \medskip \\
     -i \omega \, \varepsilon_b \, u_2 + \varepsilon_b \, \sigma_y u_2 - \dfrac{\partial v_y}{\partial y}  =  0 \medskip \\
     -i\omega \mu_b \,  \textbf{v} + \mu_b \left(\begin{array}{cc}
          \sigma_x & 0 \\
          0 & \sigma_y
     \end{array} \right)
        \textbf{v} - \nabla(u_1+u_2)  =  0\\
        u = 0 \quad \text{ at the border of the PML}.
\end{array}
\right.
$$
We consider the unknowns:
$$
u=u_1+u_2
$$
$$
u^* = u_1 - u_2.
$$
$u$, $u^*$, $v$, are solutions of the following system,

\begin{equation}
    \left\{\begin{array}{l}
         -i \omega \, \varepsilon_b \, u + \varepsilon_b \dfrac{\sigma_x+\sigma_y}{2} u + \varepsilon_b \dfrac{\sigma_x-\sigma_y}{2} u^* - \text{div} \, \textbf{v}  \, = \, 0 \medskip \\
         -i\omega \, \varepsilon_b \, u^* + \varepsilon_b \dfrac{\sigma_x+\sigma_y}{2} u^* + \varepsilon_b \dfrac{\sigma_x-\sigma_y}{2} u - \left(\dfrac{\partial v_x}{\partial x}-\dfrac{\partial v_y}{\partial y} \right) \,  = \,  0 \medskip \\
         -i\omega \mu_b \, \textbf{v} + \mu_b \, \sigma \, \textbf{v} - \nabla \, u \, = \, 0.
    \end{array}\right.
\end{equation}
The unknown $u^*$ exists only in the PML domain. In the physical domain, only unknowns $u$ and $\textbf{v}$ are present, and we solve
$$
    \left\{\begin{array}{l}
         -i \omega \, \varepsilon(\omega) \, u - \text{div} \, \textbf{v}  \, = \, -i \omega j \medskip \\
         -i\omega \mu_b \, \textbf{v} + \mu_b \, \sigma \, \textbf{v} - \nabla \, u \, = \, 0,
    \end{array}\right.
    $$
where $j$ is the source term. Of course, additional unknowns $p$ and $q$ are added in $\Omega_{res}$ to linearize the system in $\omega$.
After discretization, we will obtain :
$$
-i \omega \textbf{M}_h \textbf{U}_h + \textbf{K}_h \textbf{U}_h = \textbf{F}_h.
$$
The matrix $\textbf{M}_h$ is symmetric, while $\textbf{K}_h$ is not. 
The left eigenvector $\textbf{x}^\bot_m$ and the right eigenvector $\textbf{x}_m$ are written as: 
$$
\textbf{x}^\bot_m = \left( \begin{array}{c} \textbf{u}^\bot_m \\ \textbf{u}^{*, \bot}_m \\ \textbf{v}^\bot_m \end{array} \right), 
\quad \textbf{x}_m = \left( \begin{array}{c} \textbf{u}_m \\ \textbf{u}^*_m \\ \textbf{v}_m \end{array} \right).
$$
We have obtained the following relations ($\lambda_m = i \tilde{\omega}_m$ is the eigenvalue associated with $\textbf{x}_m$ and $\textbf{x}^\bot_m$):
$$
\textbf{u}^\bot_m = \left( 1- \dfrac{\sigma_x+\sigma_y}{2\lambda_m} \right) \textbf{u}_m
$$

$$
\textbf{u}^{*,\bot}_m = \left( \dfrac{\sigma_x-\sigma_y}{2\lambda_m} \right) \textbf{u}_m
$$
and
$$
\textbf{v}^\bot_m = \left[ \begin{array}{c}
     \dfrac{1}{\mu_b \left(-\lambda_m + \sigma_x \right)} \left( \dfrac{\partial\textbf{u}^\bot_m}{\partial x} + \dfrac{\partial\textbf{u}^{*,\bot}_m}{\partial x} \right) \\
     \dfrac{1}{\mu_b \left(-\lambda_m + \sigma_y \right)} \left( \dfrac{\partial\textbf{u}^\bot_m}{\partial y} - \dfrac{\partial\textbf{u}^{*,\bot}_m}{\partial y} \right)
\end{array}\right].
$$
The proof is given in appendix \ref{app:BiorthoPML2D}.

\subsubsection{3-D case}

In the PMLs we have: 
$$
\left \{ 
\begin{array}{l}
     -i \omega \varepsilon_b \textbf{E} + \varepsilon \textbf{T}_{2,3,1}\textbf{E} - \nabla \times \textbf{H}^* = 0  \bigskip \\
      -i \omega \mu_b \textbf{H} + \mu \textbf{T}_{2,3,1}\textbf{H} + \nabla \times \textbf{E}^* = 0 \bigskip \\
     -i \omega \textbf{E}^* + \textbf{T}_{3,1,2}\textbf{E}^* + i \omega \textbf{E} -\textbf{T}_{1,2,3}\textbf{E}= 0 \bigskip \\
     -i \omega \textbf{H}^* + \textbf{T}_{3,1,2}\textbf{H}^* + i \omega \textbf{H}-\textbf{T}_{1,2,3}\textbf{H}= 0 \bigskip \\
     \textbf{E} \times \textbf{n} = 0 \quad \mbox{at the border of the PML, } 
     \end{array}
\right. 
$$
with $\textbf{T}_{i,j,k}=\left( \begin{array}{ccc}
\sigma_i & 0 & 0\\
0 & \sigma_j & 0\\
0 & 0 & \sigma_k
\end{array}
\right)$. The unknowns $\textbf{E}^*$ and $\textbf{H}^*$ exist only in the PML domain. In
the physical domain, there are only unknowns $\textbf{E}$ and $\textbf{H}$ (supplemented by unknowns $\textbf{P}$ and $\textbf{Q}$
 in $\Omega_{res}$) that solve \eqref{eq:MaxwellSystemPQ}.
 After discretization we will obtain:
$$ -i \omega \textbf{M}_h \textbf{U}_h + \textbf{K}_h \textbf{U}_h = \textbf{F}_h.$$
The matrices $\textbf{M}_h$ and $\textbf{K}_h$ are not symmetric (see appendix \eqref{app:BiorthoPML3D}). If we note $\textbf{x}_m = \left( \textbf{E}_m, \textbf{H}_m, \textbf{E}_m^*, \textbf{H}_m^* \right)$ the right eigenvector, the left eigenvector $\textbf{x}_m^\bot$ is given as:
\begin{equation}
\textbf{x}_m^\bot = \left( \begin{array}{c}
     \textbf{E}_m^*  \\
      -\textbf{H}_m^* \medskip \\
      \left( 1+\dfrac{\textbf{T}_{2,3,1}-\textbf{T}_{3,1,2}}{-\lambda_m + \textbf{T}_{3,1,2}} \right) \varepsilon_b \textbf{E}_m \medskip \\
      -\left( 1+\dfrac{\textbf{T}_{2,3,1}-\textbf{T}_{3,1,2}}{-\lambda_m + \textbf{T}_{3,1,2}} \right)\mu_b \textbf{H}_m \medskip \\
\end{array}                   \right). 
\label{eq:LeftEigenVecPML3D}
\end{equation}
The proof is given in appendix \ref{app:BiorthoPML3D}. Straightforward computations give that $$ \langle \textbf{M}_h \textbf{x}_m, \textbf{x}_m^\bot \rangle = \int_\Omega \dfrac{\partial (\tilde{\omega}_m \, \varepsilon(\tilde{\omega}_m))}{\partial \tilde{\omega}_m} \tilde{\textbf{E}}_m \cdot \tilde{\textbf{E}}_m - \dfrac{ \partial \left(\tilde{\omega_m} \mu(\tilde{\omega}_m) \right)}{\partial \tilde{\omega_m}} \, \tilde{\textbf{H}}_m \cdot \tilde{\textbf{H}}_m d\Omega $$
with 
\begin{equation}
\varepsilon(\omega) = \varepsilon_b \dfrac{ \left(-i \omega + \textbf{T}_{2, 3, 1}\right) \left(-i \omega + \textbf{T}_{3, 1, 2} \right) } { - i \omega \left( -i \omega + \textbf{T}_{1, 2, 3} \right) }, \quad
\mu(\omega) = \mu_b \dfrac{ \left(-i \omega + \textbf{T}_{2, 3, 1}\right) \left(-i \omega + \textbf{T}_{3, 1, 2} \right) } { - i \omega \left( -i \omega + \textbf{T}_{1, 2, 3} \right) }, \quad
\label{eq:EpsMuPML}
\end{equation} 
inside the PML. We find the announced normalization \eqref{eq:Norm} in the introduction.

\subsection{Case of metals : $\omega_0 = 0$}

In section \ref{sec:Core}, the third equation of \eqref{eq:MaxwellSystemPQ} has been multiplied by $\omega_0^2/(\varepsilon_\infty \omega_p^2)$ which vanishes when $\omega_0 = 0$. But the latter case is often interesting because it occurs for metallic materials. The linear system \eqref{eq:DiscreteMaxwell} is no longer invertible because some rows of $\textbf{K}_h$ and $\textbf{M}_h$ are null. For metals, we cannot symmetrize the linear system. Therefore the calculations made in section \ref{sec:Core} are no longer valid for metals. 
However, if we consider the nonsymmetric system \eqref{eq:MaxwellSystemPQ}, 
$$
\textbf{K}= 
\left [ 
\begin{array}{cccc}
    0 &-\nabla \times &0 &0  \\
    -\nabla \times &0 &0 &0 \\
    0 & 0 & 0 & -1  \\
    \varepsilon_\infty \omega_p^2 & 0  & \omega_0^2 & -\gamma
\end{array} \right ], \ \textbf{M} =
\left[
\begin{array}{cccc}
    \varepsilon_\infty & 0 &0 &0  \\
    0 & -\mu_0 &0 &0 \\
    0 & 0 & 1 & 0  \\
    0 & 0  & 0 & -1
\end{array}
\right ],
$$
the left eigenvector $\textbf{x}_m^\bot$ is not equal to $\textbf{x}_m$, but is given as

$$\textbf{x}_m^\bot = \left[ \begin{array}{cc}
      \tilde{\textbf{E}}_m\\
      \tilde{\textbf{H}}_m \\
      \dfrac{\omega_0^2}{\varepsilon_\infty \omega^2_p} \tilde{\textbf{P}}_m \\
      \dfrac{\tilde{\textbf{Q}}_m}{\varepsilon_\infty \omega^2_p} 
\end{array}
\right ].$$
As a result, we still obtain the modal excitation coefficient \eqref{eq:FormulaAlpha} and the normalization \eqref{eq:Norm}.

\section{Numerical results}
\label{sec:Numeric}

The numerical results have been obtained with the software {\texttt Montjoie} \cite{Montjoie} for the computation of finite element matrices $\textbf{M}_h$ and $\textbf{K}_h$ given in section \ref{sec:Core}.
In this section, all the eigenvalues of the matrix $\textbf{M}_h^{-1} \textbf{K}_h$ are computed with Lapack.
We represent adimensionalized pulsations $\omega_m$ defined as
$$ \omega_m = \dfrac{\tilde{\omega}_m}{\omega_{\mbox{adim}}} $$
where 
$$ \omega_{\mbox{adim}} = \dfrac{c_0}{L_0}, \quad L_0 = 10^{-7}. $$
$c_0$ is the speed of light and $L_0$ the characteristical length (here 100nm).
All of the eigenvalues such that $|\omega_m|< 10^{-3}$ are dropped in order to remove static modes. Since the eigenvalues are complex conjugate,
 only eigenvalues (and associated eigenvectors) such that $Re(\tilde{\omega}_m) \ge 0$ are stored. The eigenvalues such that 
$\lambda_m = \sigma_i$ ($\sigma_i$ is the damping function in PMLs) are also excluded, since the auxiliary fields $\textbf{H}, \textbf{E}^*, \textbf{H}^*$
 cannot be eliminated (division by zero) for these eigenvalues. In practice, we have observed that the associated eigenvectors have null components (at machine precision)
  for the unknown $\textbf{E}_m$ and do not contribute to the field $\textbf{E}_S$.
Finally, if two pulsations $\omega_i$, $\omega_j$ are close enough (i.e. $|\omega_i-\omega_j|<10^{-6}$) they are considered degenerate. 

In this section, the three formulas \eqref{eq:FormulaAlpha} (denoted as Usual) \eqref{eq:FormuleWei} (denoted as Alternative Source) and \eqref{eq:FormuleMarseille}  (denoted as Order2) will be compared. Since the source term $\textbf{F}_h$ is null inside the PML layers, the formula \eqref{eq:AlphaDiscrete} is equal to
$$
    \alpha_m = \dfrac{1}{i(\tilde{\omega}_m-\omega)} \langle \textbf{F}_h,\textbf{x}_m\rangle,
$$
The two formulas \eqref{eq:FormulaAlpha} and \eqref{eq:FormuleWei} are implemented by taking a different source term as explained in sections \ref{sec:Core} and \ref{sec:ComparWei}. For the formula \eqref{eq:FormuleMarseille}, we did not implement matrices $\textbf{M}_h$ and $\textbf{K}_h$ introduced in section \ref{sec:ComparMarseille}, but we use the discrete equivalent of \eqref{eq:FormuleMarseille}:
$$
    \alpha_m = \dfrac{\omega}{i \tilde{\omega}_m \, (\tilde{\omega}_m-\omega)} \langle \textbf{F}_h,\textbf{x}_m\rangle,
$$
with the source term $\textbf{F}_h$ of section \ref{sec:Core}.

\subsection{2-D disk}

\begin{figure}[!h]
\centerline{\includegraphics[height=6cm]{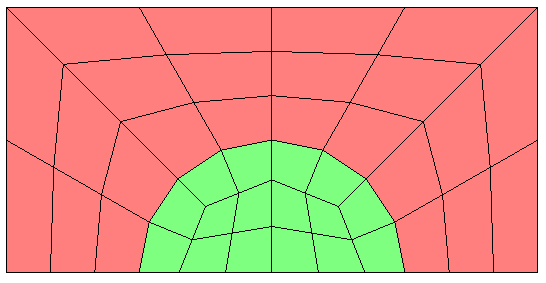}}
\caption{Mesh used for the scattering of a disk}
\label{fig:Maillage2D}
\end{figure}

\begin{figure}[!h]
\captionsetup[subfigure]{justification=centering}
\begin{subfigure}[h]{0.35 \textwidth}
\centerline{\includegraphics[height=4.2cm]{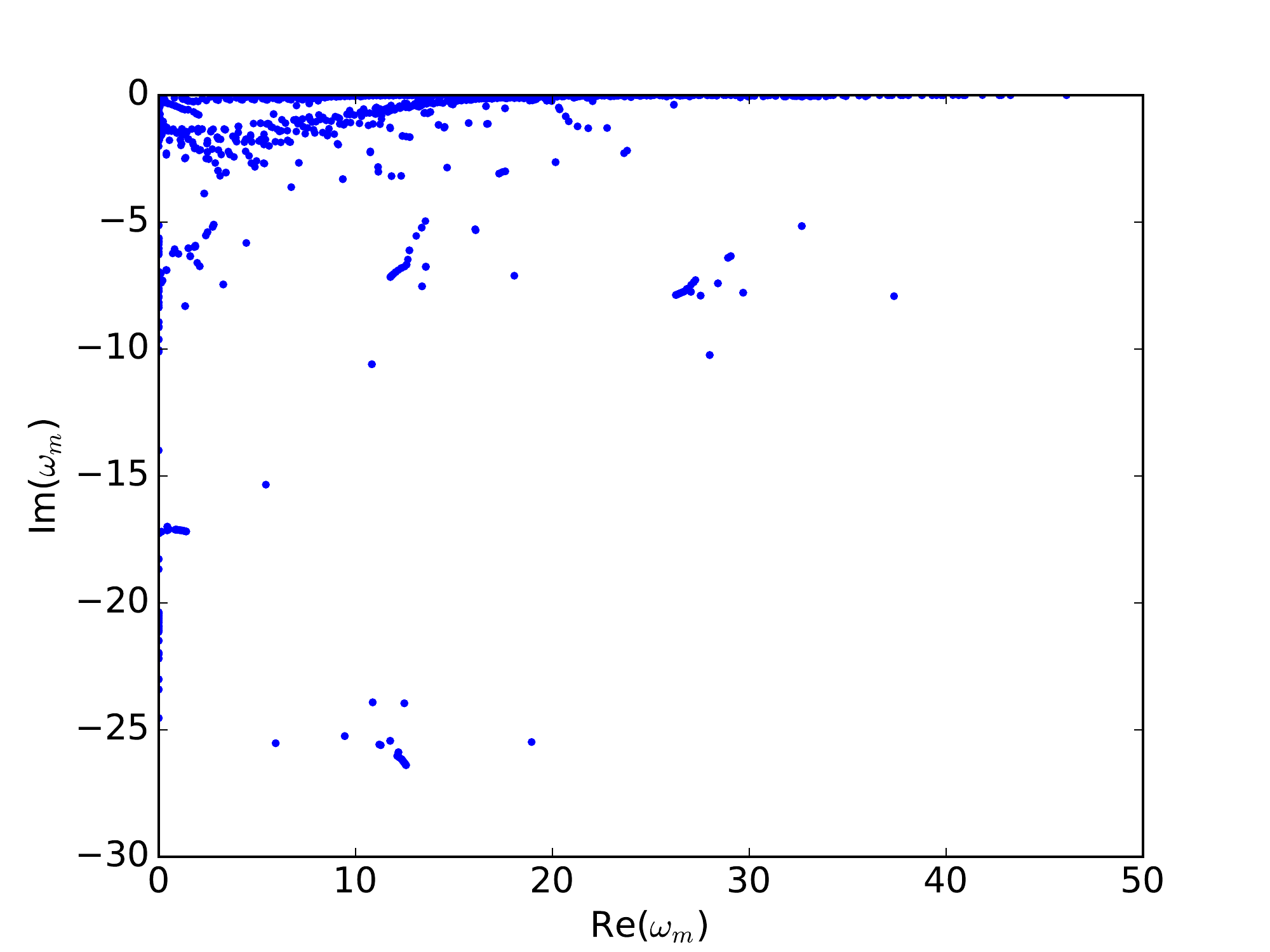}}
\caption{Whole spectrum}
\label{fig:SpectrumDisque}
\end{subfigure}
\begin{subfigure}[h]{0.65 \textwidth}
\centerline{\includegraphics[height=5cm]{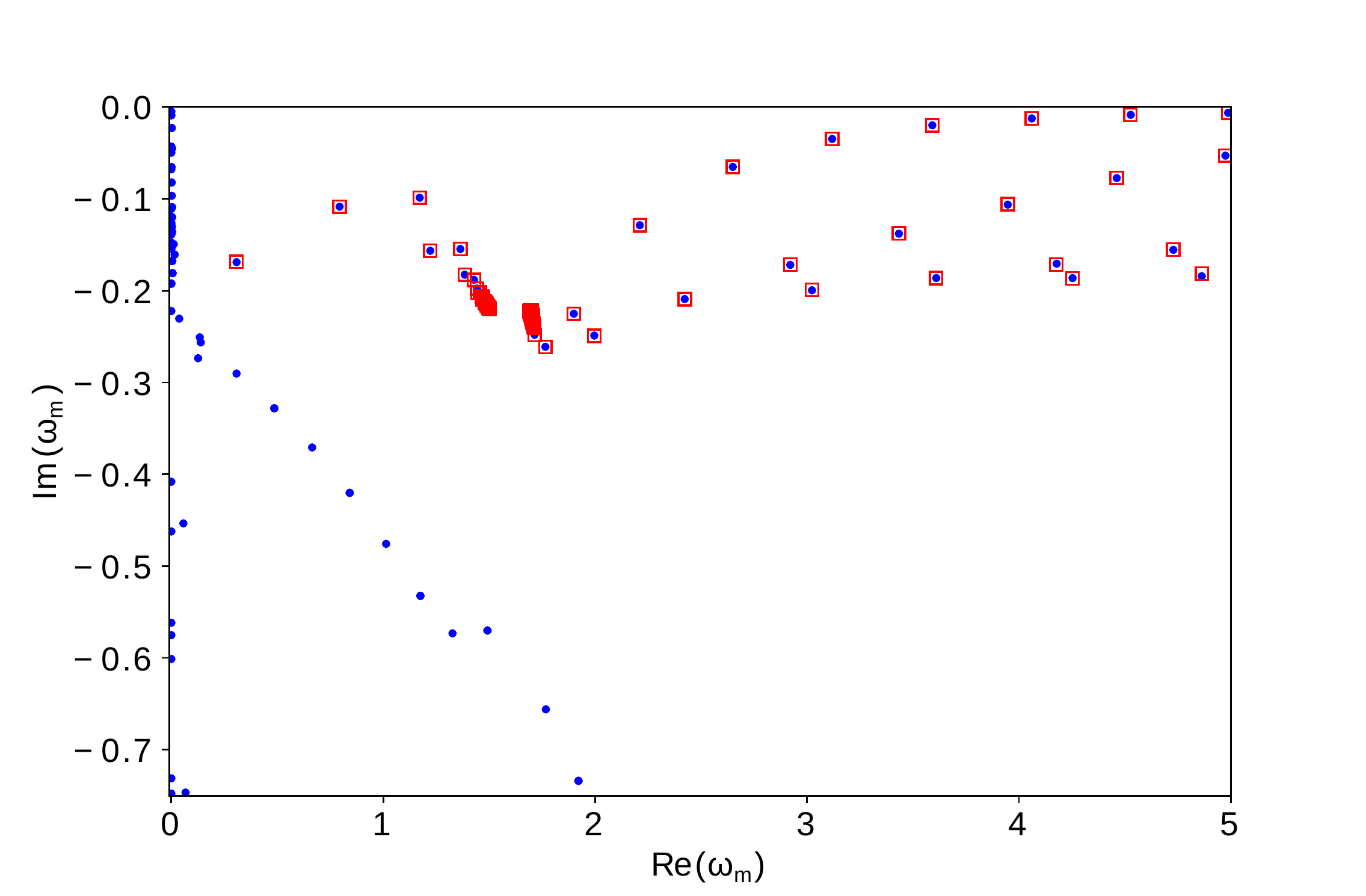}}
\caption{Part of the spectrum. \\ In red, analytical QNM pulsations.}
\label{fig:SpectrumDisqueAnal}
\end{subfigure}
\caption{Numerical adimensionalized pulsations for the disk $\omega_m$ for the disk (blue points).}
\end{figure}

We first look at the case of the field diffracted by a dielectric disk with a radius of 100 nm, where the material is modeled by a Lorentz model with 
$$ \varepsilon_\infty = 6, \quad \omega_0 = 4.572 \cdot 10^{15} \text{rad/s}, \quad \omega_p = \dfrac{\omega_0}{2}, \quad \gamma = 1.332\cdot 10^{15} \text{rad/s} $$
The physical computation domain is 400 nm long and 200 nm wide (see figure \ref{fig:Maillage2D}). PML layers are added to the mesh of figure \ref{fig:Maillage2D}. The thickness of PML is equal to 100nm with two cells in direction of PMLs. The damping of PMLs $\sigma$ is taken equal to $3$.
The field driving the system is a TE polarized plane wave, propagating along the x-axis, at the real frequency $\omega$. As a result only the component $E_z$ is non null and is discretized with continuous finite elements (here $\mathbb{Q}_4$ with the mesh of figure \ref{fig:Maillage2D}). 
\begin{figure}[!h] 
\centerline{\includegraphics[height=3.5cm]{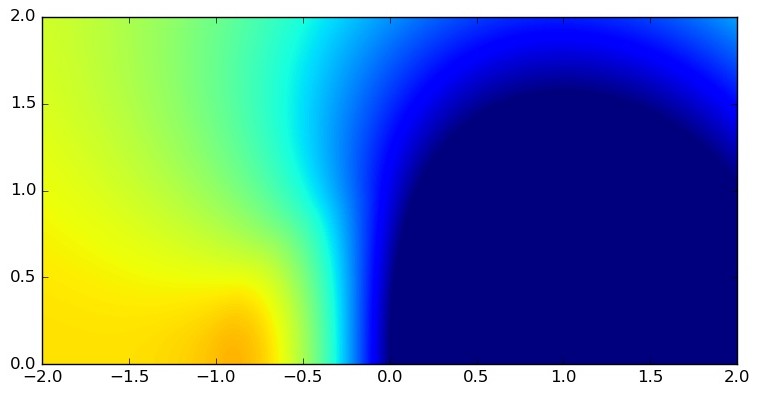} \includegraphics[height=3.5cm]{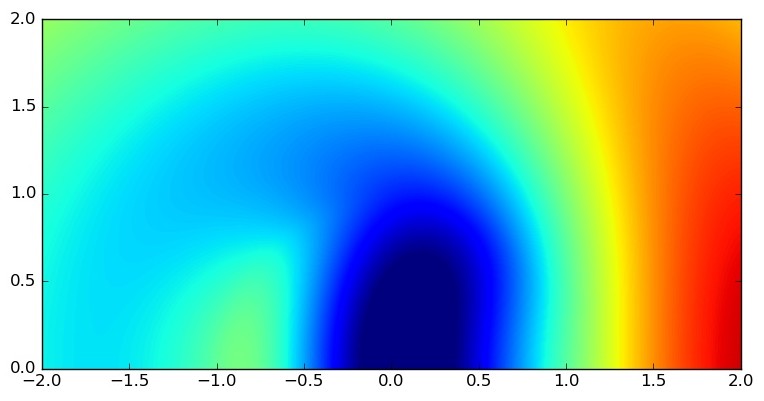}}
\centerline{\includegraphics[height=3.5cm]{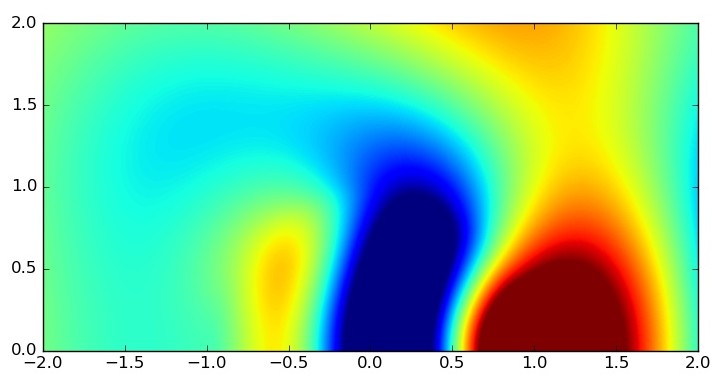} \includegraphics[height=3.5cm]{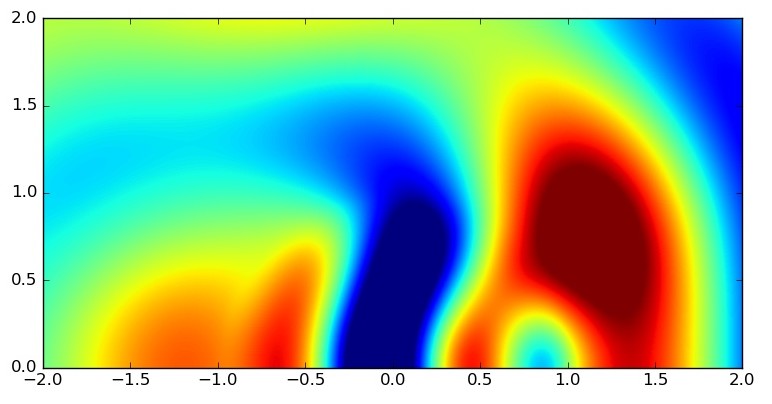}} 
\caption{Real part of the scattered field for $\omega = \omega_0/2$(top-left), $\omega = \omega_0$ (top-right), $3 \, \omega_0/2$ (bottom-left) , $2 \, \omega_0$(bottom-right).} 
\label{fig:SolQNM}
\end{figure}  
The solution is plotted for four frequencies in figure \ref{fig:SolQNM}. For the maximal frequency $\omega = 2 \omega_0$, we have computed a relative $L^2$ error of 0.164\% between the numerical solution and the analytical solution (computed with Hankel functions). We compute the field diffracted by the disk  for 31 angular frequencies $\omega$ evenly spaced in the interval $[\omega_0/2, 2 \omega_0]$. We represent in figure \ref{fig:SpectrumDisque} the adimensionalized pulsations $\omega_m$. 
We can compare these pulsations with analytical QNMs for the disk (computed with Bessel functions).
The comparison is displayed in figure \ref{fig:SpectrumDisqueAnal}. We see that QNM's are correctly computed, and the presence of other modes that we call PML modes. We observe also two accumulation points corresponding to a pole and a zero of $\varepsilon(\omega)$.
\begin{figure}[!h]
    \centering
    \includegraphics[height=8cm]{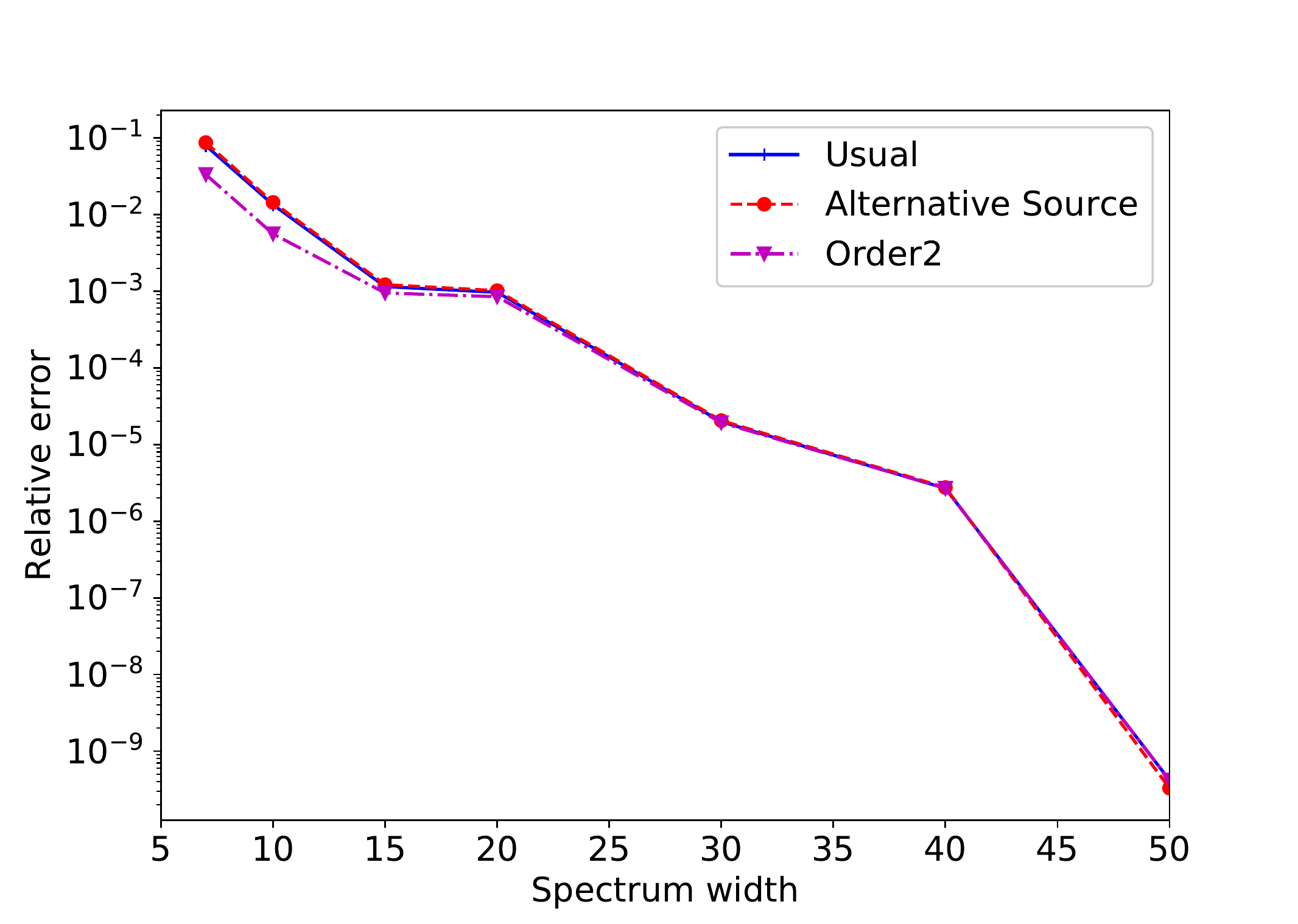}
    \caption{Relative error between the scattered field computed with the modal expansion and with a direct FEM solver as a function of the spectral width. }
    \label{fig:Convergence_TE_PML}
\end{figure}
The matrices $\textbf{M}_h$ and $\textbf{K}_h$ have 5300 rows. Among the 1798 eigenvectors stored, 286 are associated with a degenerate eigenvalue.
In figure \ref{fig:Convergence_TE_PML}, we display the relative error between the modal solution 
$$
\textbf{E}_{S}^{\mbox{modal}} = \sum_m \alpha_m \tilde{\textbf{E}}_m
$$
and the direct FEM solution 
$$
\textbf{E}_{S}^{FEM} = (-i\omega \textbf{M}_h + \textbf{K}_h)^{-1}\textbf{F}_h
$$
as a function of the width of the spectrum. For a given spectral width, the relative error is computed for 31 frequencies and the maximum value of this error is retained and plotted. For a given spectral width $L$, only the modes whose eigenfrequencies $\tilde{\omega}_m$ verify 
$$
\text{Re}(\tilde{\omega}_m) \in [-L \, \omega_{\mbox{adim}}, L \, \omega_{\mbox{adim}}] \, \text{and} \, \text{Im}(\tilde{\omega}_m) \in [- \omega_{\mbox{adim}}L/2, 0]
$$
are included in the expansion. The relative error is computed on the whole physical domain $\Omega_p$ (PMLs are not included) by the formula
$$
\text{Relative Error} = \sqrt{ \dfrac{\int_{\Omega_p}  \left|\textbf{E}_{S}^{\mbox{modal}} -\textbf{E}_{S}^{FEM} \right|^2 d\Omega_p}{\int_{\Omega_p}  \left|\textbf{E}_{S}^{FEM} \right|^2 d\Omega_p}}
$$
\begin{figure}[!h]
    \centering
    \includegraphics[height=8cm]{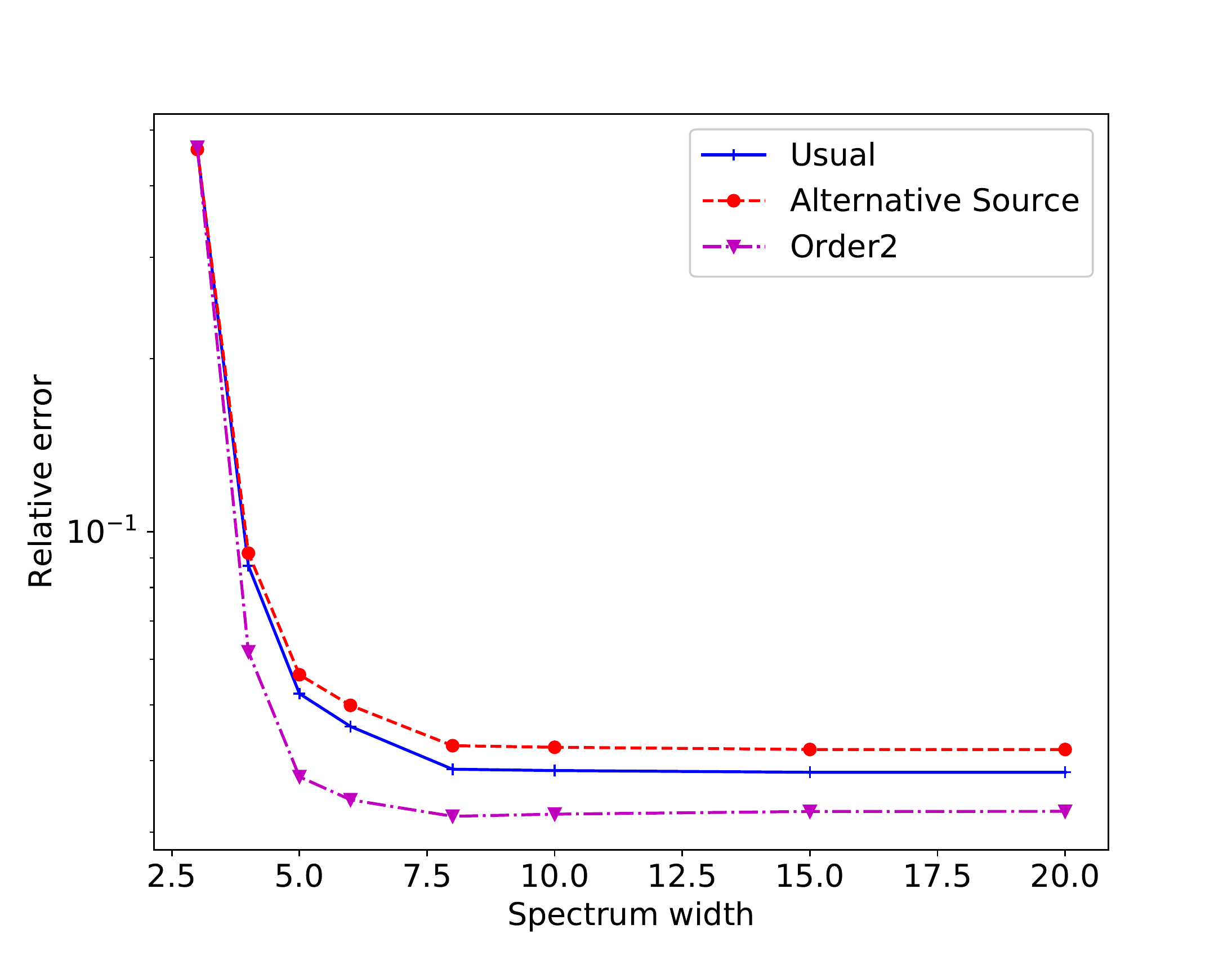}
    \caption{Relative error between the scattered field computed with the modal expansion and with a direct FEM solver as a function of the spectral width (only modes such that $Im(\omega_m) \ge -1$ are kept). }
    \label{fig:Convergence_TE_PML_Imag1}
\end{figure}
In the figure \ref{fig:Convergence_TE_PML}, the three formulas \eqref{eq:FormulaAlpha} (denoted as Usual), \eqref{eq:FormuleMarseille} (denoted as Order2) and \eqref{eq:FormuleWei} (denoted as Alternative Source) are compared. It is observed that all of these formulas provide a modal solution that converges towards the direct FEM solution as expected. The two formulas \eqref{eq:FormuleWei} and \eqref{eq:FormulaAlpha} are very close, while the last formula \eqref{eq:FormuleMarseille} is a bit more accurate when the spectral width $L$ is small. In the figure \ref{fig:Convergence_TE_PML_Imag1}, we have displayed the relative error computed on the disk (of radius 100 nm) versus the spectral width $L$, by keeping only modes satisfying
only the modes whose eigenfrequencies $\tilde{\omega}_m$ verify 
$$
\text{Re}(\tilde{\omega}_m) \in [-L \, \omega_{\mbox{adim}}, L \, \omega_{\mbox{adim}}] \, \text{and} \, \text{Im}(\tilde{\omega}_m) \in [- \omega_{\mbox{adim}}, 0]
$$
By this criterion, we tried to select mostly QNM modes, the error is computed inside the disk, since it is well-known that QNM modes are complete only in the cavity (see \cite{Leung}). As expected, we observe a stagnation of the error when $L$ grows, the formula \eqref{eq:FormuleMarseille} provides the most accurate results.

\FloatBarrier

\subsection{3-D sphere}

\begin{figure}[!h]
\centerline{\includegraphics[height=6cm]{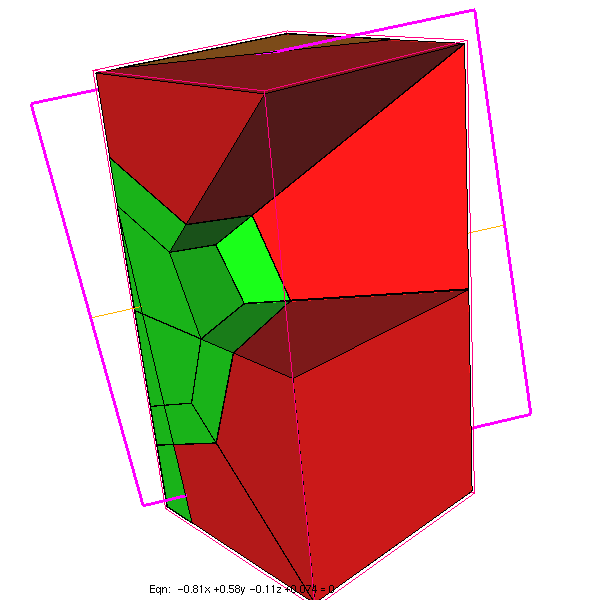}}
\caption{Mesh used for the scattering of a sphere}
\label{fig:Maillage3D}
\end{figure}

\begin{figure}[!h]
 \centerline{\includegraphics[height=8cm]{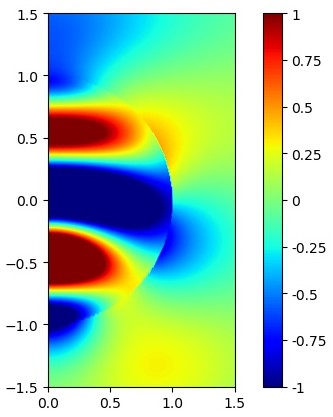}
 \includegraphics[height=8cm]{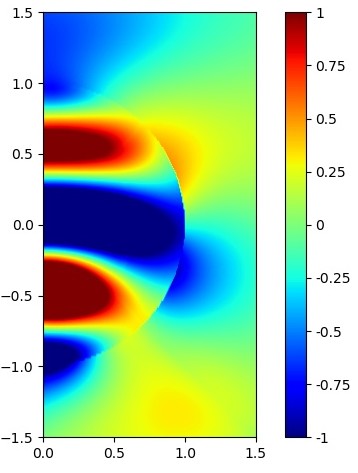} }
 \caption{Real part of diffracted field (component $E_x$ of electric field) for the plane $y=0$. On the left,
  numerical solution, on the right analytical solution. }
 \label{fig:SolSphere}
 \end{figure}

We consider the case of a field diffracted by a dielectric sphere with a radius of 100 nm with the same values as in 2-D:
$$ \varepsilon_\infty = 6, \quad \omega_0 = 4.572 \cdot 10^{15} \text{rad/s}, \quad \omega_p = \dfrac{\omega_0}{2}, \quad \gamma = 1.332\cdot 10^{15} \text{rad/s} $$
The physical computation domain is the parallepiped box $[0, 150\mbox{nm}] \times [0, 150 \mbox{nm}] \times [-150 \mbox{nm}, 150 \mbox{nm}]$ with a quarter of the dielectric ball (see figure \ref{fig:Maillage3D}). PML layers are added to the mesh of figure \ref{fig:Maillage3D}. The thickness of PML is equal to 100nm 
with only one cell in direction of PMLs. The damping of PMLs $\sigma$ is taken equal to $2$.
The source is an incident plane wave oriented in z-direction and polarized in x-direction
$$ \textbf{E}_{\mbox{inc}} = e^{i k z} \textbf{e}_x $$
We impose a Perfectly conducting condition on plane $x=0$ (i.e. $\textbf{E} \times n = 0$) and a Neumann
condition on plane $y=0$ (i.e. $\textbf{H} \times n = 0$) in order to have the same solution as for the whole sphere.
 Fourth order edge elements are used for the unknown $\mbox{E}$ and the mesh of figure \ref{fig:Maillage3D}.  We compute the field diffracted by the sphere  for 31 angular frequencies $\omega$ evenly spaced in the interval $[\omega_0/2, 2 \omega_0]$. Because of the coarse mesh, the numerical error obtained for the last frequency $2 \omega_0$ is equal to 3.73\%. This error is computed by comparing the numerical solution with the analytical solution computed with Mie's series. These two solutions are displayed in figure \ref{fig:SolSphere}.
 For this case, the matrices $\textbf{M}_h, \textbf{K}_h$ have 31 246 rows. Among the 8055 stored eigenvectors, 919 are associated with degenerate eigenvalues.
 
\begin{figure}[!h] 
 \centerline{\includegraphics[height=8cm]{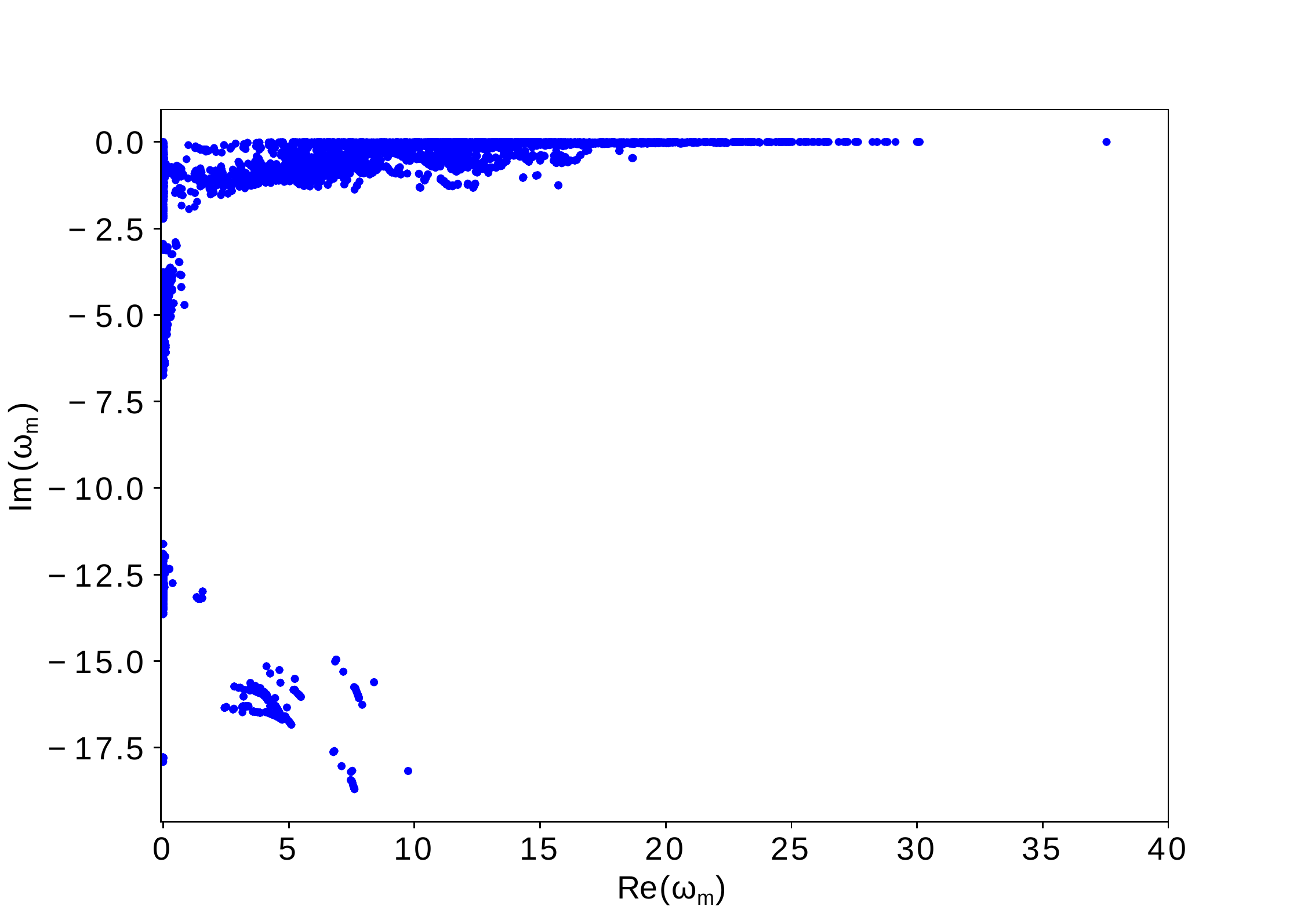}}
 \caption{Numerical adimensionalized pulsations for the sphere.}                
 \label{fig:SpectrumSphere}     
\end{figure}

\begin{figure}[!h]
\centerline{\includegraphics[height=8cm]{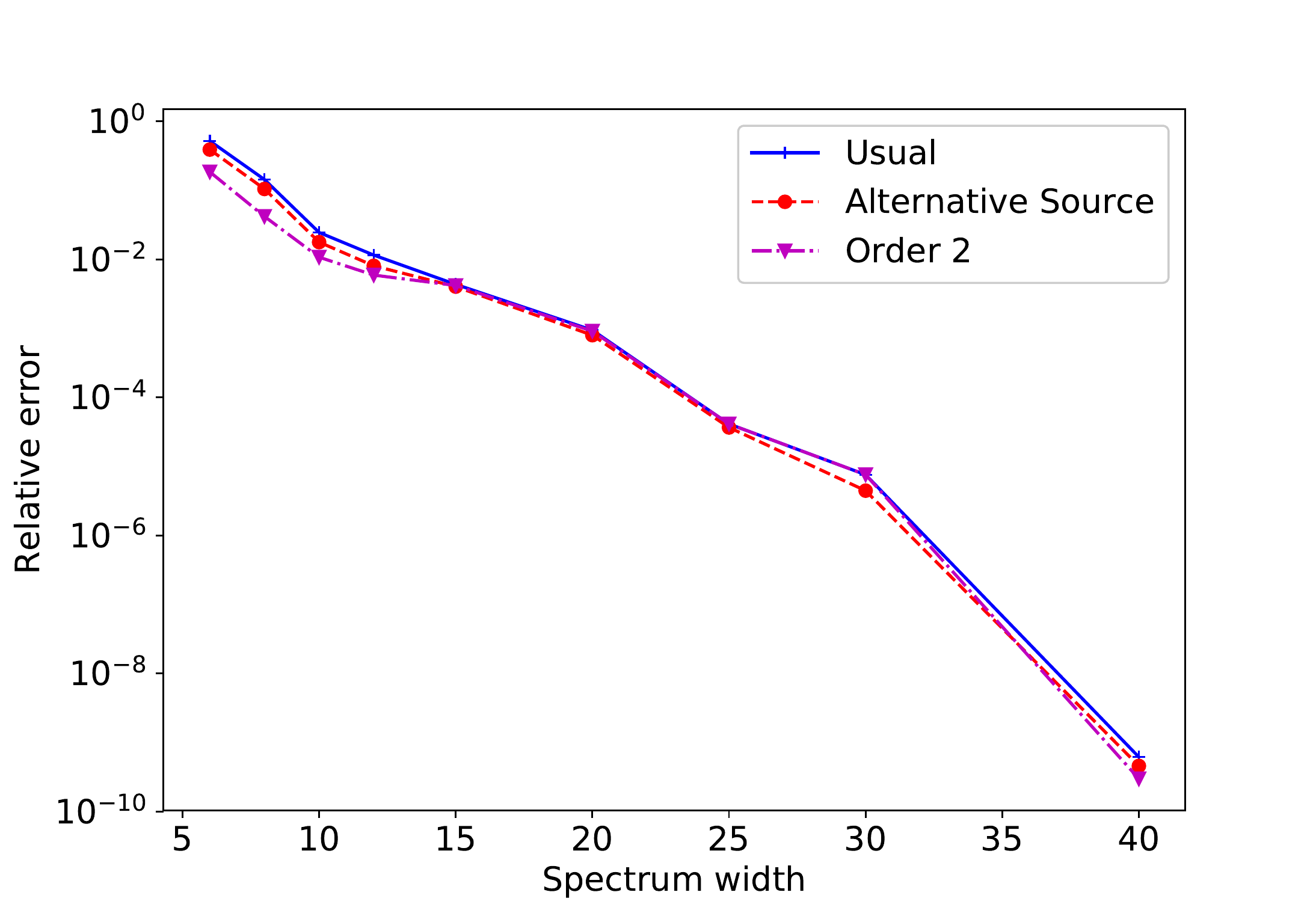}}
\caption{Relative error on curl of E versus the spectral width. Case of the sphere. }
\label{fig:ConvergenceSphere}
\end{figure}

 Numerical pulsations are plotted in figure \ref{fig:SpectrumSphere} with the same adimensionalization coefficient $\omega_{\mbox{adim}}$ as in 2-D.
 \begin{figure}[!h] 
 \centerline{\includegraphics[height=8cm]{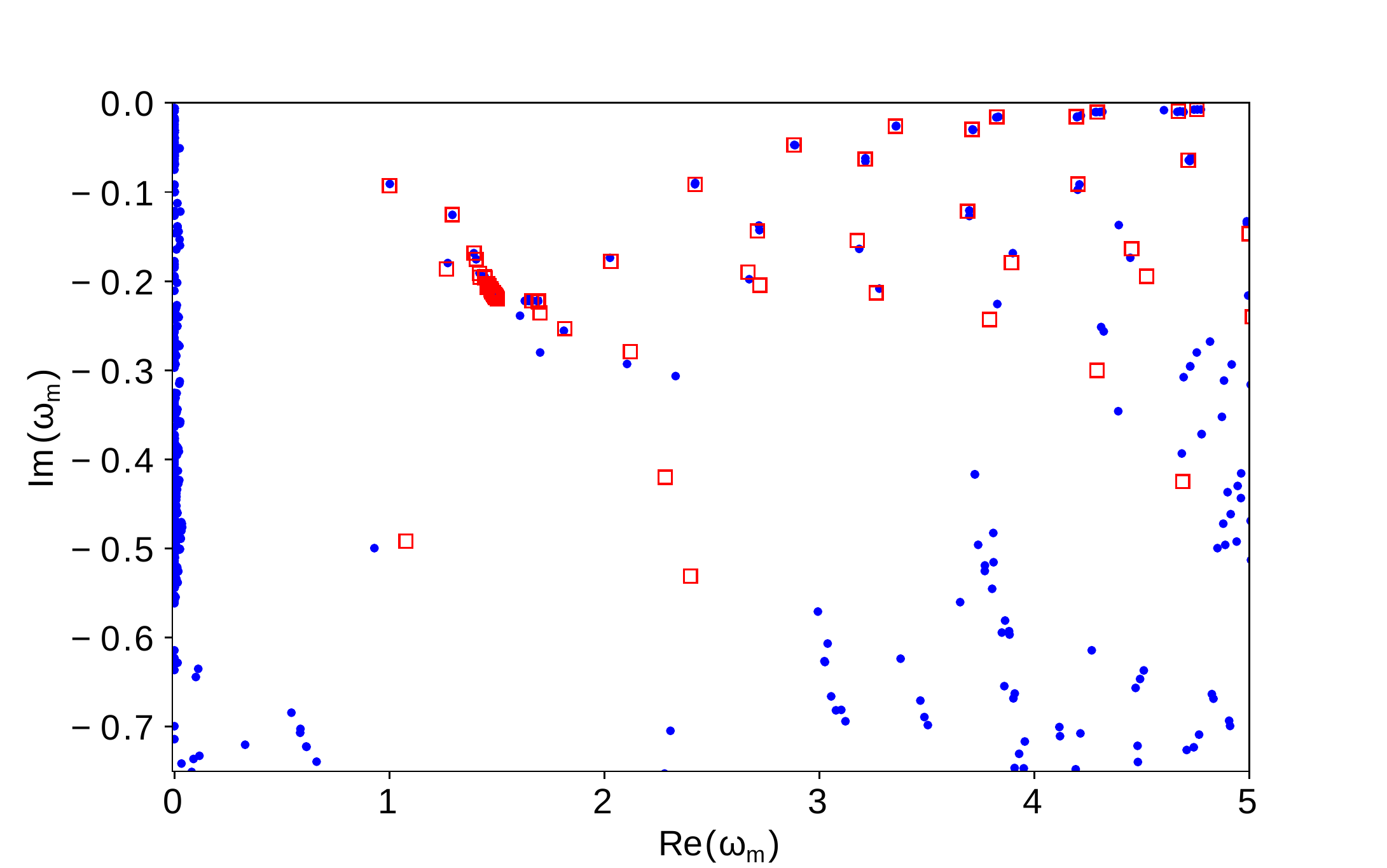}}
 \caption{Numerical adimensionalized pulsations for the sphere. Numerical eigenvalues are in blue,
  analytical QNMs in red.}                
 \label{fig:SpectrumSphereCompar}                         
\end{figure}  
When we zoom in on the box $Re(\omega) \in [0, 5 \, \omega_{\mbox{adim}}], Im(\omega) \in [-0.75 \, \omega_{\mbox{adim}}, 0.0]$, we obtain
 pulsations $\omega_m$ of the figure \ref{fig:SpectrumSphereCompar}. In this figure, we have also represented
  the analytical pulsation of QNMs. Since the mesh is much coarser in 3-D, some QNMs are not correctly approximated. We have two accumulation points, one for 
  $$ \omega/\omega_{\mbox{adim}} \approx 1.5088 - 0.2221i$$
  which corresponds to a pole of $\varepsilon(\omega)$ and one for
  $$ \omega/\omega_{\mbox{adim}} \approx 1.6905 - 0.2221i$$
  which corresponds to a zero of $\varepsilon(\omega)$. Similarly to the 2-D case, we compute the relative error between the modal solution and the direct FEM solution. However, the relative error is computed with the curl of E in order to remove the contribution of static modes:
  $$
\text{Relative Error} = \sqrt{ \dfrac{ \displaystyle \int_{\Omega_p}  \left| \nabla \times \textbf{E}_{S}^{\mbox{modal}} - \nabla \times \textbf{E}_{S}^{FEM} \right|^2 d\Omega_p}{\displaystyle \int_{\Omega_p}  \left| \nabla \times \textbf{E}_{S}^{FEM} \right|^2 d\Omega_p}}
$$
This error is plotted in figure \ref{fig:ConvergenceSphere} for formulas \eqref{eq:FormulaAlpha}, \eqref{eq:FormuleMarseille}and \eqref{eq:FormuleWei}.
Similarly to what has been observed in 2-D, the three formulas provide a modal solution that converges towards the direct FEM solution. Similarly to the 2-D case, only modes such that
$$
\text{Re}(\tilde{\omega}_m) \in [-L \, \omega_{\mbox{adim}}, L \, \omega_{\mbox{adim}}] \, \text{and} \, \text{Im}(\tilde{\omega}_m) \in [- \omega_{\mbox{adim}}L/2, 0]
$$
are kept, where $L$ is the spectral width.
When a reduced spectrum is selected, the formula \eqref{eq:FormuleMarseille} is the most accurate. If the electric field is desired, a nice approach consists in discretizing $\textbf{H}$ with edge elements (instead of $\textbf{E}$), reconstructing $\textbf{H}$ with the modal expansion:
$$ \textbf{H}^{\mbox{modal}} = \sum \alpha_m \tilde{\textbf{H}}_m. $$
and of computing $\textbf{E}$ by using Maxwell's equations
\begin{equation}
\textbf{E} = \dfrac{1}{-i \omega \varepsilon(\omega)} \left( \textbf{J} + \nabla \times \textbf{H}^{\mbox{modal}} \right) 
\label{eq:ReconstructE}
\end{equation}
\begin{figure}[!h]
\centerline{\includegraphics[height=8cm]{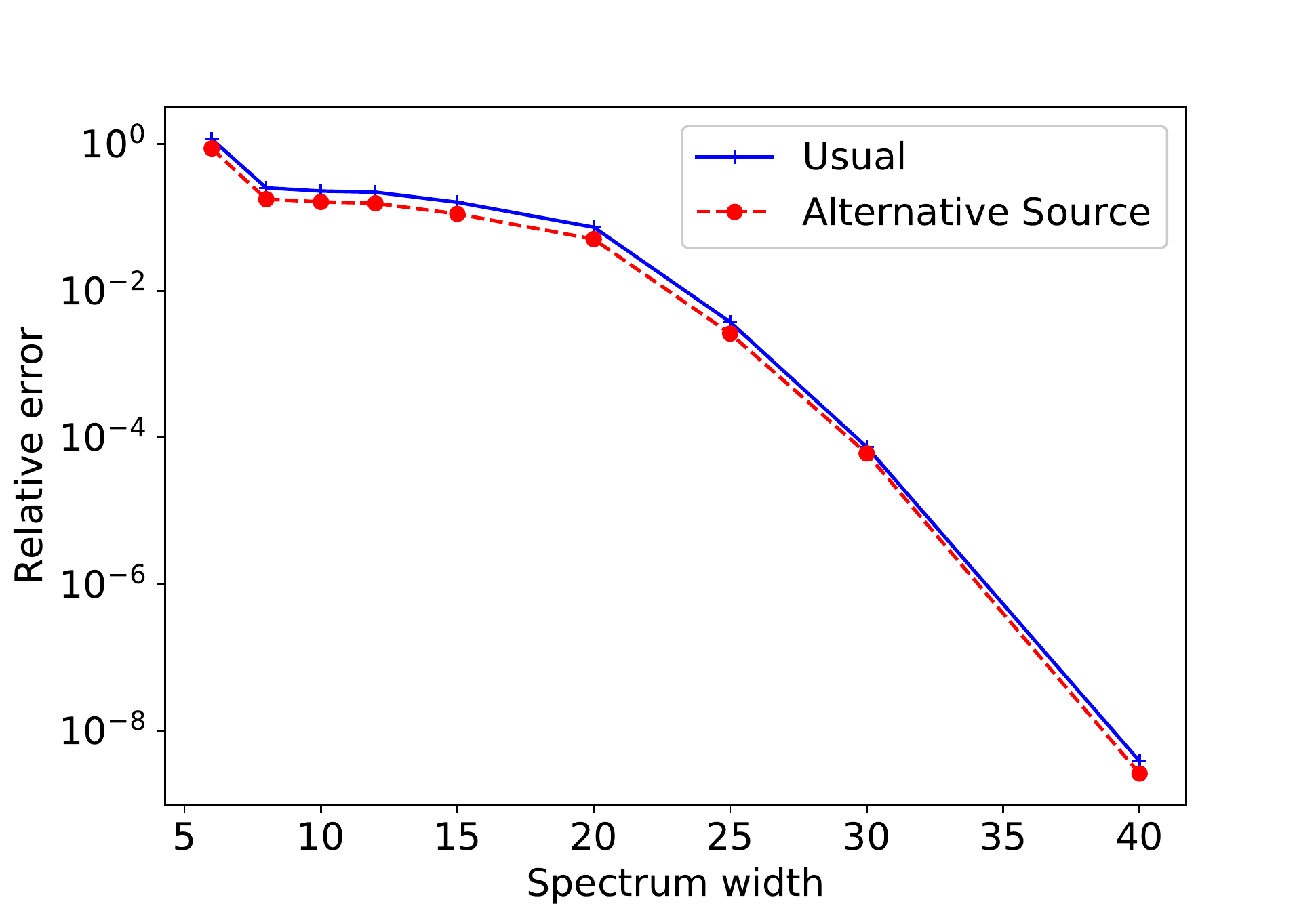}}
\caption{Relative error on electric field E (as computed in \eqref{eq:ReconstructE}) versus the spectral width. Case of the sphere. }
\label{fig:ConvergenceSphereE}
\end{figure}
In figure \ref{fig:ConvergenceSphereE}, the relative error on the electric field has been computed by using this method. Only the formulas \eqref{eq:FormulaAlpha} and \eqref{eq:FormuleWei} can be used to obtain $\textbf{H}^{\mbox{modal}}$ with the coefficients $\alpha_m$. The coefficients $\alpha_m$ given by the formula \eqref{eq:FormuleMarseille} can be used only to reconstruct $\textbf{E}^{\mbox{modal}}$ (with equation \eqref{eq:modal_expansion}). The reason is that this formula has been established by introducing the unknowns $\textbf{E}, \textbf{E}', \textbf{P}, \textbf{Q}$ (see section \eqref{sec:ComparMarseille}). Thus, only these four unknowns can be reconstructed with this formula and not $\textbf{H}$.  In figure \ref{fig:ConvergenceSphereE}, we observe that the reconstructed field $\textbf{E}$ with this method converges correctly to the numerical eletrical field. However, the accuracy obtained on $\textbf{E}$ is not as good as the accuracy we obtained on $\textbf{H}$ (in figure \ref{fig:ConvergenceSphere}).



\section{Acknowledgements}

Alexandre Gras acknowledges the support of the DGA and INRIA. Philippe Lalanne would like to thank Boris Gralak and Guillaume Demesy for fruitful discussions.

\section{Funding}

This work was funded by the Agence Nationale de la Recherche (ANR-16-CE24-0013), the Agence de l'Innovation de la D\'efense (DGA), and the Institut National de Recherche en Informatique et en Automatique (INRIA).

\section{Conclusion}

In this paper, we have discussed how the scattered field $\textbf{E}_S, \textbf{H}_S$ can be computed from the discrete eigenmodes of Maxwell's equations. Due to the discrete nature of the problem, these discrete eigenmodes form a complete basis, i.e. the numerical solution can be written exactly as a combination of the eigenmodes. However, there is no uniqueness of the coefficients $\alpha_m$ that appear in the expansion. We have shown that an infinity of formulas exists for the computation of $\alpha_m$. New formulas can be found by choosing a different linearization of dispersive Maxwell's equations or a different splitting of the source term. With our common formalism, we have been able to recover the three formulas \eqref{eq:FormulaAlpha}, \eqref{eq:FormuleMarseille} and \eqref{eq:FormuleWei} that have been previously proposed in the literature. Numerical experiments show that all these formulas converge towards the numerical solution. In the tested cases, we observed that the formula \eqref{eq:FormuleMarseille} is slightly more accurate than other formulas when a small part of the eigenvalues are selected. 
We also explain how degenerate eigenvalues are treated with a simple Gram-Schmidt orthogonalization. This procedure is essential in order to construct an orthogonal basis of eigenmodes with respect to matrix $\textbf{M}_h$, which can be seen as a non-classical scalar product. We detail how dispersive PMLs can be handled with our formalism. Because of the symmetry of the original dispersive Maxwell's equations, there is no need to compute the biorthogonal eigenvector (or left eigenvector) since this eigenvector can be computed directly from the right eigenvector. However, for more complex cases such as gratings with quasi-periodic conditions where the Maxwell's equations are no longer symmetric, the computation of left eigenvectors would be required.

\FloatBarrier

\bibliographystyle{apalike}
\bibliography{main}

\appendix

\section{Finite element matrices}
\label{app:FemMatrices}

In 3-D case, edge elements are used for the unknowns $\textbf{E}$ and discontinuous finite elements for unknown $\textbf{H}$ $\textbf{P}$ and $\textbf{Q}$ (see \cite{thesis_Ndiaye}), let us introduce the stiffness and mass matrices $\textbf{K}_h$ and $\textbf{M}_h$:
$$
\textbf{M}_h= \left[
\begin{array}{cccc}
     \textbf{D}_h^E & 0 & 0 & 0 \\
     0 & -\mu_0 \textbf{D}_h^H & 0 & 0 \\
     0 & 0 & \dfrac{\omega_0^2}{\varepsilon_\infty\omega_p^2} \textbf{D}_h & 0 \\
     0 & 0 & 0 & - \dfrac{\textbf{D}_h}{\varepsilon_\infty\omega_p^2} 
\end{array}
\right]
$$
$$
\textbf{K}_h = \left[
\begin{array}{cccc}
     0 & -\textbf{R}_h & 0 & \textbf{C}_h \\
     -\textbf{R}_h^T & 0 & 0 & 0 \\
     0 & 0 & 0 & -\dfrac{\omega_0^2 \textbf{D}_h }{\varepsilon_\infty\omega_p^2} \\
     \textbf{C}_h^T & 0 & -\dfrac{\omega_0^2 \textbf{D}_h}{\varepsilon_\infty\omega_p^2}  & - \dfrac{\gamma \textbf{D}_h}{\varepsilon_\infty\omega_p^2} 
\end{array}
\right]
$$
where $h$ denotes the mesh size and
$$(\textbf{D}_h^E)_{i,j} = \int_\Omega \varepsilon_e \, \boldsymbol{\varphi}_i(\textbf{r}) \cdot \boldsymbol{\varphi}_j(\textbf{r}) d\textbf{r} $$
$$(\textbf{D}_h)_{i,j} = \int_{\Omega_{res}} \boldsymbol{\varphi}_i(\textbf{r}) \cdot \boldsymbol{\varphi}_j(\textbf{r}) d\textbf{r} $$
$$(\textbf{D}_h^H)_{i,j} = \int_\Omega \boldsymbol{\psi}_i(\textbf{r}) \cdot \boldsymbol{\psi}_j(\textbf{r}) d\textbf{r} $$
$$(\textbf{C}_h)_{i,j} = \int_{\Omega_{res}} \boldsymbol{\varphi}_i(\textbf{r}) \cdot \boldsymbol{\psi}_j(\textbf{r}) d\textbf{r} $$
$$(\textbf{R}_h)_{i,j} = \int_\Omega \boldsymbol{\psi}_j(\textbf{r}) \cdot \nabla \times \boldsymbol{\varphi}_i(\textbf{r}) d\textbf{r} $$
where $\varphi_i$ are basis functions used for $\textbf{E}$ and $\psi_i$ basis functions for $\textbf{H}$, $\textbf{P}$ or $\textbf{Q}$. Here, we  consider
$$ \varepsilon_e =  \left \{ \begin{array}{l}
\varepsilon_\infty \, \mbox{ in } \Omega_{res} \\
\varepsilon_b, \mbox{ elsewhere. }
\end{array} \right.
 $$
 For the matrix $\textbf{R}_h$, there is no surface integral, since we will impose $\textbf{E} \times \textbf{n} = 0$ or $\textbf{H} \times \textbf{n} = 0$ on the boundaries. Degrees of freedom for $\textbf{P}$ and $\textbf{Q}$ are restricted to the domain $\Omega_{res}$. Since the matrix $\textbf{D}_h$ is symmetric, the matrices $\textbf{M}_h$ and $\textbf{K}_h$ are real symmetric. 
 
 For the 2-D case, the unknown $\textbf{E}$ is scalar (we consider the Transverse Electric case) 
  and discretized with nodal continuous elements, unknowns $\textbf{P}$ and $\textbf{Q}$ are also scalar and discretized with basis functions of $\textbf{E}$. The unknown $\textbf{H}$ is vectorial and discretized with discontinuous elements.

\section{Computation of biorthogonal vector for 2-D PML}
\label{app:BiorthoPML2D}

The proof of the relations given in subsection \ref{sec:PML2D} is done with continuous operators $\textbf{M}$ and $\textbf{K}$. Its extension to discrete operators (i.e. matrices $\textbf{M}_h$ and $\textbf{K}_h$) is straightforward thanks to mass lumping. 
We have the eigenvalue problem: 

$$
\textbf{K} \textbf{x}_m = \lambda_m \textbf{M} \textbf{x}_m
$$

where 

$$
\textbf{x}_m = \left( \begin{array}{c}
       \tilde{u}_m \\
       u^*_m \\
       \textbf{v}_m
\end{array} \right)
$$

and 
$$
\textbf{M} = \left( \begin{array}{ccc}
     \varepsilon_b & 0 & 0 \\
     0 & \varepsilon_b & 0 \\
     0 & 0 & -\mu_b
\end{array} \right) , \,
\textbf{K} = \left( \begin{array}{ccc}
     \varepsilon_b \left( \dfrac{\sigma_x + \sigma_y}{2} \right) & \varepsilon_b \left( \dfrac{\sigma_x - \sigma_y}{2} \right) & -\text{div} \medskip \\
     \varepsilon_b \left( \dfrac{\sigma_x - \sigma_y}{2} \right) & \varepsilon_b \left( \dfrac{\sigma_x + \sigma_y}{2} \right) & -\text{div}_\bot \medskip \\
     \nabla & 0 & -\mu_b \, \sigma \medskip
\end{array} \right)
$$

where $\text{div}_\bot v = \dfrac{\partial v_x}{\partial x} - \dfrac{\partial v_y}{\partial y}$
and $\nabla_\bot = \left| \begin{array}{c} \dfrac{\partial}{\partial x} \medskip \\ - \dfrac{\partial}{\partial y}\end{array}  \right.$

We want to know $\textbf{x}_m^\bot$, the eigenvector of the adjoint problem: 

$$
\textbf{K}^T \textbf{x}_m^\bot = \lambda_m \textbf{M}^T \textbf{x}_m^\bot
$$

where $\textbf{x}_m^\bot$ is split into three components : 
$$
\textbf{x}_m^\bot = \left( \begin{array}{c} u_m^\bot \\ u^{*,\bot}_m \\ \textbf{v}_m^\bot \end{array} \right)
$$

We write the matrix $\textbf{K}^T$ of the adjoint problem.
$$
\textbf{K}^T = \left( \begin{array}{ccc}
     \varepsilon_b \left( \dfrac{\sigma_x + \sigma_y}{2} \right) & \varepsilon_b \left( \dfrac{\sigma_x - \sigma_y}{2} \right) & -\text{div} \medskip \\
     \varepsilon_b \left( \dfrac{\sigma_x - \sigma_y}{2} \right) & \varepsilon_b \left( \dfrac{\sigma_x + \sigma_y}{2} \right) & 0 \medskip \\
     \nabla & \nabla_\bot & -\mu_b \, \sigma
\end{array} \right)
$$
The second equation of the adjoint problem yields a relation between $u_m^\bot$ and $u_m^{*,\bot}$:
$$
\left( \dfrac{\sigma_x - \sigma_y}{2} \right) u_m^\bot + \left( \dfrac{\sigma_x + \sigma_y}{2} \right) u_m^{*,\bot} = \lambda_m u_m^{*,\bot} $$
We infer that 
$$ u_m^{*,\bot} = \dfrac{ (\sigma_x - \sigma_y) \tilde{u} }{ 2\lambda_m - (\sigma_x + \sigma_y) }
$$
The third equation gives us $\textbf{v}_m^\bot$ as a function of $u_m^\bot$ and $u_m^{*,\bot}$
$$\textbf{v}_m^\bot = \left[ \mu_b \left(-\lambda_m + \sigma\right) \right]^{-1} \left( \nabla u_m^\bot + \nabla_\bot u_m^{*,\bot} \right) $$

with the first equation being
$$
\varepsilon_b \left( \dfrac{\sigma_x + \sigma_y}{2} \right) u_m^\bot + \varepsilon_b \left( \dfrac{\sigma_x - \sigma_y}{2} \right)u_m^{*,\bot} - \text{div} \, \textbf{v}_m^\bot = \lambda_m \varepsilon_b u_m^\bot
$$

Using the two previous equations, we now get
$$
\begin{array}{l}
\varepsilon_b \left( \dfrac{\sigma_x + \sigma_y}{2} - \lambda_m \right)u_m^\bot + \varepsilon_b \left( \dfrac{\sigma_x - \sigma_y}{2} \right)^2 \dfrac{u_m^\bot}{\lambda_m - \dfrac{\sigma_x + \sigma_y}{2}} \\
\qquad \; - \; \text{div} \left( \mu_b^{-1} (-\lambda_m + \sigma)^{-1} \left(\nabla u_m^\bot + \nabla_\bot \left(\dfrac{ (\sigma_x - \sigma_y) u_m^\bot }{ 2\lambda - (\sigma_x + \sigma_y)}\right) \right) \right) = 0 
\end{array}
$$

and we denote $y$ as 
$$
y = - \text{div} \left( \mu_b^{-1} (-\lambda_m + \sigma)^{-1} \left(\nabla u_m^\bot + \nabla_\bot \left( \dfrac{ (\sigma_x - \sigma_y) u_m^\bot }{ 2\lambda_m - (\sigma_x + \sigma_y)} \right) \right) \right)
$$
The part of the variational formulation associated with $y$ will provide
$$
\begin{array}{lll}
\displaystyle \int_\Omega \, y \, \varphi \, d\Omega  & = & \displaystyle \int_\Omega \dfrac{\mu_b^{-1}}{-\lambda_m + \sigma_x} \dfrac{\partial u_m^\bot}{\partial x} \dfrac{\partial \varphi}{\partial x} + \dfrac{\mu_b^{-1}}{-\lambda_m + \sigma_y} \dfrac{\partial u_m^\bot}{\partial y} \dfrac{\partial \varphi}{\partial y}  \medskip \\
 & & \displaystyle + \dfrac{\mu_b^{-1}}{-\lambda_m + \sigma_x} \dfrac{\partial}{\partial x} \left( \dfrac{\sigma_x - \sigma_y}{2\lambda_m - (\sigma_x + \sigma_y)} u_m^\bot \right) - \dfrac{\mu_b^{-1}}{-\lambda_m + \sigma_y} \dfrac{\partial}{\partial y}\left( \dfrac{\sigma_x - \sigma_y}{2\lambda_m - (\sigma_x + \sigma_y)} u_m^\bot \right)
  \end{array}
$$
Since the damping $\sigma_y$ does not depend on $x$, we have
$$
\dfrac{\partial}{\partial x} \left( \dfrac{2\lambda_m - (\sigma_x + \sigma_y) + (\sigma_x - \sigma_y)}{2\lambda_m-(\sigma_x+\sigma_y)}u_m^\bot \right) = 
2(\lambda_m - \sigma_y) \dfrac{\partial}{\partial x} \left( \dfrac{u_m^\bot}{2\lambda_m - \sigma_x + \sigma_y} \right) = 2(\lambda_m - \sigma_y) \dfrac{\partial u_m}{\partial x} ,
$$
with 
$$u_m = \dfrac{u_m^\bot}{-\lambda_m + \left( \dfrac{\sigma_x + \sigma_y}{2} \right)}.$$
Similarly, we proove
$$ 
\dfrac{\partial}{\partial y} \left( \dfrac{2\lambda_m - (\sigma_x + \sigma_y) - (\sigma_x - \sigma_y)}{2\lambda-(\sigma_x+\sigma_y)}u_m^\bot \right) = 2(\lambda_m - \sigma_x) \dfrac{\partial u_m}{\partial y} .
$$
As a result, we obtain
$$
\int_\Omega \, y \, \varphi \, d\Omega = \int_\Omega  \mu_b^{-1} \, \left( \dfrac{-\lambda_m + \sigma_y}{-\lambda_m + \sigma_x} \right) \dfrac{\partial u_m}{\partial x} \dfrac{\partial \varphi}{\partial x} + \mu_b^{-1} \left( \dfrac{-\lambda_m + \sigma_x}{-\lambda_m + \sigma_y} \right) \dfrac{\partial u_m}{\partial y} \dfrac{\partial \varphi}{\partial y} 
 \, d\Omega $$

For the mass terms, we have
$$
\varepsilon_b \left[ \left( -\lambda_m + \dfrac{\sigma_x + \sigma_y}{2} \right)^2 - \left( \dfrac{\sigma_x - \sigma_y}{2} \right)^2 \right] u_m = \varepsilon_b (-\lambda_m + \sigma_x)(-\lambda_m + \sigma_y) u_m
$$
Therefore, the unknown $u_m$ satisfies the following variational formulation
$$
\int_\Omega \, \varepsilon_b \, (-\lambda + \sigma_x) \, (-\lambda+\sigma_y) \, u_m \, \varphi  + \mu_b^{-1} \left( \begin{array}{cc} \displaystyle \dfrac{-\lambda + \sigma_y}{-\lambda + \sigma_x} & 0 \medskip \\
  0 & \displaystyle \dfrac{-\lambda + \sigma_x}{-\lambda + \sigma_y}  \end{array} \right) \nabla u_m \cdot \nabla \varphi \, d\Omega = 0
$$
which is the same variational formulation satisfied by $\tilde{u}_m$ (component of the eigenvector of $\textbf{K} \textbf{x}_m = \lambda_m \textbf{M} \textbf{x}_m$). $u_m$ is proportional to $\tilde{u}_m$ if $\lambda_m$ is a simple eigenvalue. In order to have $u_m^\bot = \tilde{u}_m$ in the physical domain, we will divide by $-\lambda_m$. We therefore have the following relation
$$
u_m^\bot = \left( 1 - \dfrac{\sigma_x + \sigma_y}{2 \lambda_m} \right) \tilde{u}_m 
$$
We infer that
$$
u_m^{*,\bot} = \dfrac{\sigma_x - \sigma_y}{2 \lambda_m} \tilde{u}_m
$$
For the last component $\textbf{v}_m^\bot$, it is computed from $u_m^\bot$ and $u_m^{*,\bot}$
$$
\textbf{v}_m^\bot = \left| \begin{array}{c}
     \dfrac{\mu_b^{-1}}{-\lambda_m+\sigma_x}\left( \dfrac{\partial u_m^\bot}{\partial x} + \dfrac{\partial u_m^{*,\bot}}{\partial x} \right)\\
     \dfrac{\mu_b^{-1}}{-\lambda_m+\sigma_y}\left( \dfrac{\partial u_m^\bot}{\partial y} - \dfrac{\partial u_m^{*,\bot}}{\partial y} \right)\\      
\end{array} \right.
$$

\section{Computation of biorthogonal vector for 3-D PML}
\label{app:BiorthoPML3D}

We have the eigenvalue problem
$$ \textbf{K} \textbf{U} = \lambda_m \textbf{M} \textbf{U} $$
where $\lambda_m = i \tilde{\omega}_m$ is the eigenvalue with  $$\textbf{M} = \left( \begin{array}{cccc}
     \varepsilon_b & 0 & 0 & 0 \\
     0 &  \mu_b & 0 & 0 \\
     -1 & 0 & 1 & 0 \\
     0 & -1 & 0 & 1 
\end{array}\right) , \;
\textbf{K} = \left( \begin{array}{cccc}
     \varepsilon_b \textbf{T}_{2,3,1} & 0 & 0 & -\nabla \times \\
     0 &  \mu_b  \textbf{T}_{2,3,1} & \nabla \times & 0 \\
     -\textbf{T}_{1,2,3} & 0 & \textbf{T}_{3,1,2} & 0 \\
     0 & -\textbf{T}_{1,2,3} & 0 & \textbf{T}_{3,1,2} 
\end{array}\right), \; \textbf{U} = \left( \begin{array}{c}
\textbf{E} \\
\textbf{H} \\
\textbf{E}^* \\
\textbf{H}^*
\end{array}\right)$$

In order to find the left eigenvector of this system, we consider the adjoint eigenvalue problem to this system. 

$$
\textbf{K}^T \textbf{U}^\bot = \lambda \textbf{M}^T \textbf{U}^\bot ,
$$
with $\textbf{U}^\bot = \left(\textbf{E}^\bot, \textbf{H}^\bot, \textbf{E}^{*,\bot}, \textbf{H}^{*,\bot} \right)$,
which grants us the following system of equations:
\begin{equation}
\left \{
\begin{array}{l}
     \varepsilon_b (-\lambda_m + \textbf{T}_{2,3,1})\textbf{E}^\bot - (-\lambda_m + \textbf{T}_{1,2,3})\textbf{E}^{*,\bot} = 0 \smallskip \\
     \mu_b (-\lambda_m + \textbf{T}_{2,3,1}) \textbf{H}^\bot -(-\lambda_m + \textbf{T}_{1,2,3})\textbf{H}^{*,\bot} = 0 \smallskip \\
     (-\lambda_m + \textbf{T}_{3,1,2})\textbf{E}^{*,\bot} + \nabla \times \textbf{H}^\bot = 0 \smallskip \\
     (-\lambda_m+\textbf{T}_{3,1,2})\textbf{H}^{*,\bot} - \nabla \times \textbf{E}^\bot = 0 \smallskip
     \end{array}
\right.
\label{eq:BiorthoPML3D}
\end{equation}
We are now going to try to identify the different components of $\textbf{U}^\bot$. 

First off we can show $\textbf{E}^\bot = \textbf{E}^*$. The third equation and second equation of \eqref{eq:BiorthoPML3D} give:
$$
\textbf{E}^{*,\bot} = \dfrac{-\nabla \times \textbf{H}^\bot}{-\lambda_m+\textbf{T}_{3,1,2}} = \dfrac{-1}{-\lambda_m+\textbf{T}_{3,1,2}}\nabla \times \left( \dfrac{-\lambda_m+\textbf{T}_{1,2,3}}{\mu_b (-\lambda_m+\textbf{T}_{2,3,1})} \textbf{H}^{*,\bot} \right) 
$$

The first equation and fourth equation of \eqref{eq:BiorthoPML3D}, provide
$$ \textbf{E}^{*,\bot}  =  \dfrac{\varepsilon_b (-\lambda_m + \textbf{T}_{2,3,1})}{-\lambda + \textbf{T}_{1,2,3}} \textbf{E}^\bot, \quad \textbf{H}^{*, \bot} = \dfrac{ \nabla \times \textbf{E}^\bot}{-\lambda_m + \textbf{T}_{3,1,2}}$$
By substituting these expressions in the previous equation, we obtain an equation in $\textbf{E}^\bot$ only:
$$
\dfrac{\varepsilon_b (-\lambda_m + \textbf{T}_{2,3,1})}{-\lambda_m + \textbf{T}_{1,2,3}} \textbf{E}^\bot  = \dfrac{-1}{-\lambda_m+\textbf{T}_{3,1,2}}\nabla \times \left(  - \dfrac{-\lambda_m+\textbf{T}_{1,2,3}}{\mu_b (-\lambda_m+\textbf{T}_{2,3,1})(-\lambda_m + \textbf{T}_{3,1,2})} \nabla \times \textbf{E}^\bot \right) 
$$
Since $\textbf{E}^*$ verifies the same eigenvalue problem, we can choose the constant such that
$$ \textbf{E}^\bot = \textbf{E}^* $$
Next, we will show that $\textbf{H}^\bot = - \textbf{H}^*$. 

Using $ \textbf{H}^{*,\bot} = \dfrac{\nabla \times \textbf{E}^*}{-\lambda_m + \textbf{T}_{3,1,2}}$ and 
$ \textbf{H}^{*,\bot} = \dfrac{\mu_b(-\lambda_m + \textbf{T}_{2,3,1})}{-\lambda_m + \textbf{T}_{1,2,3}} \textbf{H}^\bot$, we can show that 

$$\textbf{H}^\bot = \dfrac{-\lambda_m+\textbf{T}_{1,2,3}}{\mu_b (-\lambda_m+\textbf{T}_{2,3,1})(-\lambda_m+\textbf{T}_{3,1,2})} \nabla \times \textbf{E}^* = - \textbf{H}^*$$

and from there, it can easily be shown that

$$
\textbf{E}^{*,\bot} = \dfrac{-\lambda_m+\textbf{T}_{2,3,1}}{-\lambda_m+\textbf{T}_{3,1,2}} \, \varepsilon_b \, \textbf{E}
$$

$$
\textbf{H}^{*,\bot} =- \dfrac{-\lambda_m+\textbf{T}_{2,3,1}}{-\lambda_m+\textbf{T}_{3,1,2}}\, \mu_b \, \textbf{H}
$$

Which we can rewrite: 

$$
\textbf{E}^{*,\bot} = \left( 1+\dfrac{\textbf{T}_{2,3,1}-\textbf{T}_{3,1,2}}{-\lambda_m+\textbf{T}_{3,1,2}} \right)\varepsilon\textbf{E} ,\quad \textbf{H}^{*,\bot} = -\left( 1+\dfrac{\textbf{T}_{2,3,1}-\textbf{T}_{3,1,2}}{-\lambda_m+\textbf{T}_{3,1,2}} \right)\mu\textbf{H} , 
$$
Therefore we have obtained the left eigenvector given in formula \eqref{eq:LeftEigenVecPML3D}.

\end{document}